\documentclass[12pt,journal,onecolumn,letterpaper]{IEEEtran}

\newcommand{\bfLambda}{\mathbf{\Lambda}}
\newcommand{\bftildeQ}{\mathbf{\tilde{Q}}}

\newcommand{\bftildeY}{\mathbf{\tilde{Y}}}
\newcommand{\bfbreveX}{\mathbf{\breve{X}}}
\newcommand{\bfbreveQ}{\mathbf{\breve{Q}}}
\newcommand{\bfhatLambda}{\mathbf{\hat{\Lambda}}}
\newcommand\bfhat[1]{\mathbf{\hat{#1}}}
\newcommand\bftilde[1]{\mathbf{\tilde{#1}}}
\newcommand{\argmin}{\operatornamewithlimits{argmin}}

\renewcommand\baselinestretch{1.3}

\usepackage{amsmath,amssymb,graphics,epsfig,latexsym,color}
\usepackage{times}
\usepackage{subfigure}
\usepackage{psfrag}

\newtheorem{theorem}{Theorem}[section]

\newtheorem{defn}[theorem]{Definition} 
\newtheorem{lemma}[theorem]{Lemma}

\newcommand{\junk}[1]{}

\newcommand{\hbf}{\mbox{${\bf h }$} }

\newcommand{\xbf}{\mbox{${\bf x }$} }
\newcommand{\ybf}{\mbox{${\bf y }$} }
\newcommand{\zbf}{\mbox{${\bf z }$} }

\newcommand{\Qbf}{\mbox{${\bf Q }$} }

\newcommand{\Sbf}{\mbox{${\bf S }$} }

\newcommand{\Xbf}{\mbox{${\bf X }$} }
\newcommand{\Ybf}{\mbox{${\bf Y }$} }

\newcommand{\Zbf}{\mbox{${\bf Z }$} }

\newcommand{\Complex}{{\rm {\bf C}\mkern-9mu\rule{0.05em}{1.4ex}\mkern10mu}}

\begin{document}

\title{On successive refinement of diversity  for fading ISI channels}

\author{S. Dusad and S. N. Diggavi% <-this % stops a space 
  \thanks{EPFL, Lausanne, Switzerland, S. Dusad was supported in part by SNSF
  Grant \# 200021-105640/1.  S. N. Diggavi is part of the SNSF
  supported NCCR-MICS center on wireless sensor networks.  Email:
  suhas.diggavi@epfl.ch, sanket.dusad@gmail.com.  This work has appeared in part
at Allerton 2006 \cite{allerton} and ISIT 2008 \cite{DusDig08b}.} }

%\IEEEpeerreviewmaketitle
\maketitle

\begin{abstract}

Rate and diversity impose a fundamental trade-off in
communications. This trade-off was investigated for flat-fading
channels in \cite{ZhengTse03} as well as for Inter-symbol Interference
(ISI) channels in \cite{dtse}.  A different point of view was explored
in \cite{dimacs} where high-rate codes were designed so that they have
a high-diversity code embedded within them. These diversity embedded
codes were investigated for flat fading channels both from an
information-theoretic viewpoint \cite{DigTse05} and from a coding
theory viewpoint in \cite{DigCal08}. In this paper we explore the use
of diversity embedded codes for inter-symbol interference channels. In
particular the main result of this paper is that the diversity
multiplexing trade-off for fading MISO/SIMO/SISO ISI channels is indeed
successively refinable. This implies that for fading ISI channels with
a single degree of freedom one can embed a high diversity code within
a high rate code without any performance loss (asymptotically). This
is related to a deterministic structural observation about the
asymptotic behavior of frequency response of channel with respect to
fading strength of time domain taps as well as a coding scheme to take
advantage of this observation.

\end{abstract}

%%%%%%%%%%%%%%%%%%%%%%%%%%%%%%%%%%%%%%%%%%%%%%%%%%%%%%%%%%%%%%%%%%%%%%%%%%%%%%%%

\section{Introduction}
\label{sec:intro}

The classical approach towards code design for channels is to maximize
the data rate given a desired level of reliability. The classical
outage formulation divides the set of channel realizations into an
outage set $\mathcal{O}$ and a non-outage set
$\overline{\mathcal{O}}$: it requires that a code has to be designed
such that the transmitted message can be decoded with arbitrary small
error probability on all the channels in the non-outage set. Since the
code must work for all such channels, the data rate is limited by the
worst channel in the non-outage set. Note that in this scenario, the
communication strategy cannot take advantage of the opportunity when
the channel happens to be stronger than the worst channel in the
non-outage set.  For this classical approach, a seminal result in
\cite{ZhengTse03} showed that there exists a fundamental trade-off
between diversity (error probability) and multiplexing (rate). This
was characterized in the high SNR regime for flat fading channels with
multiple transmit and multiple receive antennas (MIMO)
\cite{ZhengTse03}. This D-M trade-off has been extended to several
cases including scalar (SISO) fading ISI channels \cite{dtse}. The
presence of ISI gives significant improvement of the diversity
order. In fact, for the SISO case the improvement was equivalent to
having multiple receive antennas equal to the number of ISI taps
\cite{dtse}.

\begin{figure}
     \centering
       \input{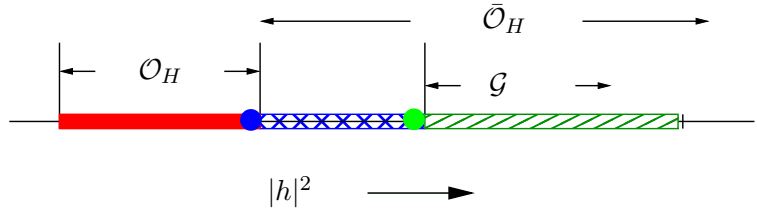}
       \caption{Outage events in the classical setting and for diversity
       embedded coding} \label{fig:DivEmb}
\end{figure}

Diversity embedded coding takes advantage of the good channel
realizations by an opportunistic coding strategy \cite{DigAld04}.
Although the focus is on two levels of diversity, the results can be
easily generalized to arbitrary number of levels. Consider two
information streams with $\mathcal{H}$ denoting the message set from
the first information stream and $\mathcal{L}$ denoting that from the
second information stream. Diversity embedded codes encode the streams
such that the high-priority stream ($\mathcal{H}$) is decoded with
arbitrary small error probability whenever the channel is not in
outage ($\mathcal{O}$) and in addition the lower-priority stream is
decoded whenever the channel is in a set $\mathcal{G}\subset
\overline{ \mathcal{O}}_H$ of good channels (see Figure
\ref{fig:DivEmb}).

In this paper we explore the performance of diversity embedded codes
\cite{DigTse05} over ISI channels with single degree of freedom {\em
  i.e.}, $min(M_t,M_r)=1$. The rates for the higher and lower priority
message sets, as a function of $SNR$, are respectively $R_H(SNR)$ and
$R_L(SNR)$. Consider transmission over a channel for which
$(r,D^{opt}(r))$ is the optimal single-layer diversity-multiplexing
point corresponding to the channel. After transmission the decoder
jointly decodes the two message sets and we can define two error
probabilities, $P_e^H(SNR)$ and $P_e^L(SNR)$, which denote the average
error probabilities for message sets $\mathcal{H}$ and $\mathcal{L}$
respectively. We want to characterize the tuple $(r_H, D_H, r_L, D_L)$
of rates and diversities channel that are achievable where,
\begin{align*}
D_H =\lim_{SNR\rightarrow\infty}
-\frac{\log {P}_e^H(SNR)}{\log(SNR)},& \,\,  r_H
=\lim_{SNR\rightarrow\infty}
\frac{R_H(SNR)}{\log(SNR)} \\
D_L =\lim_{SNR\rightarrow\infty}
-\frac{\log {P}_e^L(SNR)}{\log(SNR)},& \,\,  r_L
=\lim_{SNR\rightarrow\infty}
\frac{R_L(SNR)}{\log(SNR)}.
\end{align*}
If viewed as a single-layer code, the diversity embedded code achieves
rate-diversity pairs $(r_H,D_H)$ and $(r_H+r_L,D_L)$, where it is
assumed that $D_H\geq D_L$. Since it is not possible to beat the
single-layer D-M trade-off, note that necessarily $D_H\leq
D^{opt}(r_H)$ and $D_L\leq D^{opt}(r_H+r_L)$.

\begin{figure}
  \centering
  \includegraphics[scale=0.5]{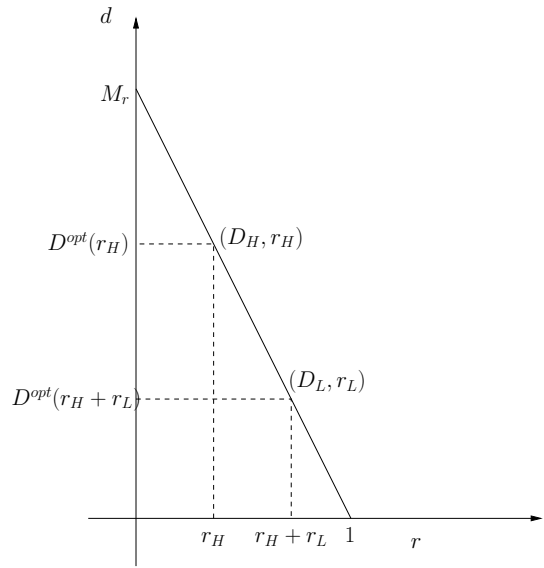}
  \caption[Successive refinement for a flat fading channel]{Successive
    refinement for a flat fading channel with $M_r$ receive antennas
    and one transmit antenna.}\label{fig:SucRef}
\end{figure}

In \cite{DigTse06} it was shown that when we have one degree of freedom
(one transmit many receive or one receive many transmit antennas) the
D-M trade-off was successively refinable. That is, the high priority
scheme (with higher diversity order) can attain the optimal
diversity-multiplexing (D-M) performance as if the low priority stream
was absent. This property of successive refinement is illustrated in Figure 
\ref{fig:SucRef}. However, the low priority scheme (with lower diversity
order) attains the same D-M performance as that of the aggregate rate
of the two streams.  When there is more than one degree of freedom
(for example, parallel fading channels) such a successive refinement
property does not hold \cite{DigTse06}.

Since the Fourier basis is the eigenbasis for linear time invariant
channels we can decompose the transmission into a set of parallel
channels. Since it is known that the D-M trade-off for parallel fading
channels is not successively refinable \cite{DigTse06}, it is tempting
to expect the same for fading ISI channels.  The main result in this
paper demonstrates that for fading ISI channels with one degree of
freedom (SISO/SIMO/MISO) the D-M trade-off is indeed successively
refinable. At first this result might seem surprising, but the
correlations of the fading across the parallel channels cause the
difference in the behavior.

For the SISO ISI case we show that uncoded transmission is sufficient
to demonstrate successive refinability. For the MISO case we need to
develop a coding strategy related to universal codes \cite{tavildar}
to obtain our main result. Surprisingly like the flat fading case, the
ISI fading channel with single degree of freedom (SISO/SIMO/MISO) is
successively refinable. The main result of this paper is stated below.

\begin{theorem}
\label{thm:MainRes}
The diversity multiplexing trade-off for a $\nu$ tap point to point
MISO/SIMO/SISO ISI channel is successively refinable, {\em i.e.}, for
any multiplexing gains $r_H$ and $r_L$ such that $r_H+r_L \leq
\frac{T_s-\nu}{T_s}$ the achievable diversity orders given by
$D_{H}(r_H)$ and $D_{L}(r_L)$ are bounded as,
\begin{align}
(\nu+1)M_t\left(1-\frac{T_s}{(T_s-\nu)}r_H\right)  & \leq  D_{H}(r_H) \leq  (\nu+1)M_t\left(1-{r_H}\right), \\
(\nu+1)M_t\left(1-\frac{T_s}{(T_s-\nu)}(r_{H}+r_L)\right) & \leq   D_{L}(r_{L}) \leq  (\nu+1)M_t\left(1-(r_H+r_L)\right)
\end{align}
where $T_s$ is finite and does not grow with SNR. 
\end{theorem}

Note that Theorem \ref{thm:MainRes} holds for arbitrary number of
levels of diversity, {\em i.e.}, the diversity multiplexing trade-off is
{\em infinitely divisible}. An implication of Theorem
\ref{thm:MainRes} is that for MISO/SIMO/SISO fading ISI channels, one
can design "ideal" opportunistic codes which adjust to the rate
supported by the fading channel without apriori knowing about the the
channel. This property could be used for allowing new networking
functionalities through opportunistic scheduling \cite{DusDig08} as
well as wireless multimedia delivery. In summary, we believe that this
property and the code construction used to achieve this result could
be important for future broadband wireless system design.

The paper is organized as follows. In Section \ref{sec:Problem} we
formulate the problem statement and present the notation.  A crucial
structural observation on the behavior of ISI fading channels is
established in \ref{chap5:sec:StrObs}.  Using these observations we
show the successive refinability of SISO and SIMO ISI channels using
uncoded QAM codes in Section \ref{chap5:sec:SIMTra}. In Section
\ref{chap5:sec:MISTra}, we propose a transmission technique to code
across space-time-frequency suitable for fading MISO ISI channels; for
such codes with a non-vanishing determinant criterion, we establish
the result for MISO ISI channel. We conclude with a short discussion
in Section \ref{sec:Disc}. Some of the more detailed proofs are
provided in the Appendices.

\section{Problem Statement}
\label{sec:Problem} 

Our focus is on the quasi-static fading ISI channel where we transmit
information coded over $M_t$ transmit antennas with $M_r$ antennas at
the receiver. Throughout this paper, we assume that the transmitter
has no channel state information (CSI), whereas the receiver is able
to perfectly track the channel (a common assumption, see for example
\cite{BigPro98,TarSes98}).

The coding scheme is limited to one quasi-static transmission block of
large enough block size $T\geq T_{thr}$ to be specified later. The
received vector at time $n$ after demodulation and sampling can be
written as
\begin{equation}
  \label{eq:ISIMIMOmodel}
        {\bf y}[n] = {\bf H}_0 {\bf x}[n] + {\bf H}_1 {\bf x}[n-1] +\ldots
        +{\bf H}_{\nu}{\bf x}[n-\nu] + {\bf z}[n]
\end{equation}
where ${\bf y} \in \Complex^{M_r\times 1}$ is the received vector at
time $n$, ${\bf H}_l \in \mathbb{C}^{M_r \times M_t}$ represents the
$l^{th}$ matrix tap of the MIMO ISI channel, ${\bf x}[n] \in
\Complex^{M_t\times 1}$ is the space-time coded transmission vector at
time $n$ with transmit power constraint $P$ and ${\bf z} \in
\Complex^{M_r\times 1}$ is assumed to be additive white (temporally
and spatially) Gaussian noise with variance $\sigma^2$. We use $SNR$
to represent the signal to noise ratio for the period of
communication. The matrix ${\bf H}_l$ consists of fading coefficients
$h_{ij}$ which are i.i.d. $\mathcal{C}\mathcal{N}(0,1)$ and fixed for
the duration of the block length ($T$). Let $h_{i}^{(p,q)}$ represent
the $i^{th}$ tap coefficient between the $p^{th}$ receive antenna and
the $q^{th}$ transmit antenna, $x^{(q)}[k]$ and $y^{(p)}[n]$ be the
symbol transmitted on the $q^{th}$ transmit antenna and the symbol
received at the $p^{th}$ receive antenna in the $n^{th}$ time instant,
respectively.

Also, let $\xbf^{(q)}_{[a,b]}$ and $y^{(p)}_{[a,b]}$ represent the
symbols transmitted on the $q^{th}$ transmit antenna and received at
the $p^{th}$ receive antenna over the time period $a$ to $b$, {\em
  i.e.},
\begin{align*}
%  \mathbf{y}^{(p)}_{[0,T_s-1]} & =  \left[ \begin{array}{c} y^{(p)}[0] \\ y^{(p)}[1] \\ \vdots \\ y^{(p)}[T_s-1]  \end{array} \right].
  \mathbf{y}^{(p)}_{[0,T_s-1]} & =  \left[ \begin{array}{cccc} y^{(p)}[0] & y^{(p)}[1] & \ldots & y^{(p)}[T_s-1]  \end{array} \right]^t.
\end{align*}

Consider a sequence of coding schemes with transmission rate as a
function of $SNR$ given by $R(SNR)$ and an average error probability
of decoding ${P}_e(SNR)$. Analogous to \cite{ZhengTse03} we define
the multiplexing rate $r$ and the diversity order $D$ as follows,
\begin{equation}
\label{eq:DivTuple} D =\lim_{SNR\rightarrow\infty}
-\frac{\log {P}_e(SNR)}{\log(SNR)} ,\,\, r
=\lim_{SNR\rightarrow\infty}
\frac{R(SNR)}{\log(SNR)}.
\end{equation}
We use the special symbol $\doteq$ to denote exponential equality {\em
  i.e.}, we write $f(SNR)\doteq SNR^b$ to denote
\begin{align*}
\lim_{SNR\rightarrow\infty}\frac{\log f (SNR)}{\log(SNR)} & =  b 
\end{align*}
and $\stackrel{\cdot}{\leq}$ and $\stackrel{\cdot}{\geq}$ are defined similarly. We use the following definition for successive refinability.
\begin{defn} \cite{DigTse04}
\label{thm:SucRef}
A channel is said to be successively refinable if the
diversity-multiplexing trade-off curve for transmission is successively
refinable, {\em i.e.,} for any multiplexing gains $r_H$ and $r_L$ such
that $r_H+r_L \leq min(M_t,M_r) $, the diversity orders
\begin{align}
\label{eq:SucRef}
D_H  = D^{opt}(r_H),& \,\,D_L = D^{opt}(r_H+r_L)
\end{align}
are achievable, where $D^{opt}(r)$ is the optimal diversity order of the
channel.
\end{defn}

The concept of successive refinability can be visualized as in Figure
\ref{fig:SucRef}. For codes that are successively refinable this
definition implies that one can perfectly embed a high diversity code
within a high rate code {\em i.e.}, the high-priority can attain the
optimal diversity performance as though the low-priority stream were
not there and yet the diversity performance of the low priority stream
is the same as the optimal diversity of a stream with the aggregate
rate of the two streams.

From an information-theoretic point of view \cite{DigTse05} focused on
the case when there is one degree of freedom, ({\em i.e.,}
$\min(M_t,M_r)=1$), and transmission over a flat fading Rayleigh
channel. In that case if we consider $D_H \geq D_L$ without loss of
generality, it was established \cite{DigTse05} that the channel is successively 
refinable.  This implies that for channels with a single degree of freedom
$\min(M_t,M_r)=1$, we can design ideal {\em opportunistic} codes and
that the D-M trade-off for SIMO/MISO are successively refinable. The
question of successive refinability was further investigated in
\cite{DigTse06} for $K$ parallel i.i.d channels, the simplest of the
channels with multiple degrees of freedom, and it was shown that the
channel is not successively refinable. In particular, if we desire the 
optimal performance for the higher layer stream ($r_H$) then there is a 
loss of diversity of $(K-1)r_H$ due to the embedding and therefore the $K$
parallel i.i.d. channel is not successively refinable.

The diversity multiplexing trade-off for a {\em scalar} fading ISI
channel was established in \cite{dtse}, and the result is
summarized below.
\begin{theorem}
\label{thm:ISIDivMulTraRep}
\cite{dtse} The diversity multiplexing trade-off for transmission
over a SISO ISI channel with $\nu+1$ taps for transmission over a
period of time $T_s$ assuming perfect channel knowledge only at the
receiver for $0\leq r \leq \frac{T_s-\nu}{T_s}$  is bounded by
\begin{equation}
\label{eq:ISIDMtradeoff}
(\nu+1)\left(1-\frac{T_s}{T_s-\nu}r \right) \leq D_{isi}(r) \leq  (\nu+1)\left(1-r\right).
\end{equation}
\end{theorem}

The D-M trade-off for the SIMO channel can also be easily obtained
using techniques similar to the proof of this result. Since the D-M trade-off for parallel
independent channels is not successively refinable and given the
derivation of the SISO ISI trade-off it might be tempting to conclude
that the D-M trade-off for the ISI channel is not successively
refinable. We will show in this paper that the ISI channel is
successively refinable by utilizing the fact that correlations exist
across these sets of independent parallel channels.

\section{Structural Observation}
\label{chap5:sec:StrObs}
In this section we make a deterministic structural observation
relating the value of the taps in frequency domain to the value of the
taps in time domain.  To make the observation we consider the MIMO
model in (\ref{eq:ISIMIMOmodel}) and consider the specific
transmission scheme as in the previous section where we transmit for a
period of $T_s-\nu$ time instants and pad it with $\nu$ zero symbols. We
refer to this zero padded block of length $T_s$ as one symbol. The
received symbols over the period of $T_s$ can be written as,

{\footnotesize
\begin{align}
\underbrace{
\left[\begin{array}{c} \mathbf{y}[0]\\  \mathbf{y} [1] \\ \vdots \\ \mathbf{y}[T_s-\nu-1]  \\  \ldots \\ \mathbf{y}[T_s-1] \end{array} \right]
}_\mathbf{Y}
 & =
\underbrace{
\left[
\begin{array}{cccccccc}
\mathbf{H}_{0}    & \mathbf{0}      & \ldots     & \mathbf{0}      &   \mathbf{H}_{\nu} & \ldots           & \mathbf{H}_{2}  & \mathbf{H}_{1}    \\
\mathbf{H}_{1}    & \mathbf{H}_{0}  & \ldots     & \mathbf{0}      &  \mathbf{0}        & \mathbf{H}_{\nu} & \ldots          & \mathbf{H}_{2}    \\
\vdots            & \vdots          &            &                 &  \ldots            & \ldots           & \mathbf{0}      & \mathbf{0}        \\
\mathbf{0}        & \ldots          & \mathbf{0} & \mathbf{H}_{\nu}& \mathbf{H}_{\nu-1} & \ldots           &  \mathbf{H}_{1} & \mathbf{H}_{0}
\end{array}
\right]
}_{\mathbf{H}}
\underbrace{
\left[ \begin{array}{c}
\xbf[0] \\ \xbf[1] \\ \vdots \\ \xbf[T_s-\nu-1] \\ \mathbf{0}_{\nu\times 1}
\end{array}
\right]
}_{\mathbf{W}}
+ \nonumber \\
& \hspace{1in}
\underbrace{
\left[ \begin{array}{cccccc}
\mathbf{z}[0] & \mathbf{z}[1] & \ldots & \mathbf{z}[T_s-\nu-1]  & \ldots & \mathbf{z}[T_s-1]
\end{array}
\right]
}_\mathbf{Z}\label{eq:CircChan2}
\end{align}
}
%{\em i.e.},
%\begin{align}
%\label{eq:CircChan2}
%\mathbf{Y} & = \mathbf{H}\mathbf{W}+\mathbf{Z}
%\end{align}
where ${\bf Y} \in \mathbb{C}^{T_s M_r \times 1 } $, ${\bf H}\in
\mathbb{C}^{T_s M_r \times T_s M_t }$, ${\bf W}\in \mathbb{C}^{T_s M_t\times
  1 }$, ${\bf Z}\in \mathbb{C}^{T_s M_r \times 1}$. Denote
$\mathbf{C}=circ\{c_0,c_1,\ldots,c_{T_s-1}\}$ to be the $T_s \times T_s$
circulant matrix given by
\begin{align}
\label{eq:DefCirc}
\mathbf{C} & =  \left[
\begin{array}{cccccc}
c_0 & c_1 & c_2 & \ldots & c_{T_s-2} & c_{T_s-1} \\
c_{T_s-1} & c_0 & c_1 & \ldots & c_{T_s-3} & c_{T_s-2} \\
\vdots & & \vdots & \ddots &  &\vdots \\
c_1 & c_2 & c_3 & \ldots &  c_{T_s-1}  & c_0 \\
\end{array}
\right]
\end{align}
Rearranging and permuting the rows and columns of equation
(\ref{eq:CircChan2}), we get
\begin{align}
  \label{eq:CircChan3}
  \left[ \begin{array}{c} \mathbf{y}^{(1)}_{[0,T_s-1]}
      \\ \mathbf{y}^{(2)}_{[0,T_s-1]} \\\vdots
      \\ \mathbf{y}^{(M_r)}_{[0,T_s-1]} \end{array} \right] & =
  \underbrace{
    \left[
      \begin{array}{cccc}
        \mathbf{H}^{(1,1)} & \mathbf{H}^{(1,2)} & \ldots & \mathbf{H}^{(1,M_t)} \\
        \mathbf{H}^{(2,1)} & \mathbf{H}^{(2,2)} & \ldots & \mathbf{H}^{(2,M_t)} \\
        \vdots &  & \vdots & \\
        \mathbf{H}^{(M_r,1)} & \mathbf{H}^{(2,2)} & \ldots & \mathbf{H}^{(M_r,M_t)}
      \end{array}
      \right]
  }_{\mathbf{H}_{circ}}
    \left[\begin{array}{c}
      \mathbf{x}^{(1)}_{[0,T_s-1]} \\  \mathbf{x}^{(2)}_{[0,T_s-1]} \\ \vdots \\  \mathbf{x}^{(M_t)}_{[0,T_s-1]}  \end{array}\right]+
  \mathbf{Z}
\end{align}
where $\mathbf{H}^{(p,q)}$ are circulant matrices given by
\begin{align}
\label{eq:DefCircH}
  \mathbf{H}^{(p,q)} & =  circ\{h_0^{(p,q)},0,\ldots,0,h_{\nu}^{(p,q)} ,\ldots,h_2^{(p,q)},h_1^{(p,q)}  \}.
\end{align}
Since the $\mathbf{H}^{(p,q)}$ are circulant matrices they can be
written using the frequency-domain notation as
$\mathbf{H}^{(p,q)}=\mathbf{Q}\mathbf{\Lambda}^{(p,q)} \mathbf{Q^*}$
where $\mathbf{\Lambda}^{(p,q)}$ are diagonal matrices with elements
given by
\begin{align}
\label{eq:DefLambda2}
\mathbf{\Lambda}^{(p,q)} & =  diag\left\{\lambda_{k}^{(p,q)} : \lambda_{k}^{(p,q)}=\sum_{l=0}^{\nu}h_l^{(p,q)} e^{-\frac{2 \pi j}{T_s}k l}  \right\} \quad \mbox{for }  k =\{0,\ldots,(T_s-1)\}.
\end{align}

To get an intuition of the result consider the polynomial
\begin{align}
\lambda^{(p,q)}(z)  & =   \sum_{m=0}^{\nu} h_m^{(p,q)} z^{m},
\nonumber
\end{align}
which evaluates to the $k^{th}$ tap coefficient in the frequency
domain for $z=e^{-\frac{2 \pi j}{T_s}k}$. Since this is a polynomial
of maximum degree $\nu$, if we evaluate the polynomial at
$z=e^{-\frac{2 \pi j}{T_s}k}$ for $k =\{0,\ldots,(T_s-1)\}$, at most
$\nu$ values can be zero and at least $T_s-\nu$ values are bounded
away from zero. The following lemma formalizes this intuition and relates the asymptotic behaviors of the
frequency-domain coefficients to the fading strength of
the time-domain taps. This lemma is then used in the remaining
sections to show the successive refinement of the ISI trade-off.
\begin{lemma}
\label{lem:TapMag} Consider the taps in the frequency domain in
(\ref{eq:DefLambda2}) given by
\begin{align*}
\lambda_{k}^{(p,q)} & =  \sum_{l=0}^{\nu}h_l^{(p,q)} e^{-\frac{2 \pi j}{T_s}k l}
\end{align*}
for $k =\{0,\ldots,(T_s-1)\}$, $\ p \in\{1,\ldots,M_r\}$ and $\ q
\in\{1,\ldots,M_t\}$. For $\alpha \in (0,1]$, define the sets
  $\mathcal{G}^{(p,q)}$, $\mathcal{F}^{(p,q)}(\alpha)$ and
  $\mathcal{M}(\alpha)$ as
\begin{align}
\mathcal{G}^{(p,q)} & =  \{k : |\lambda_k^{(p,q)}|^2 \doteq \max_{l\in\{ 0,1,\ldots,\nu\} }|h_l^{(p,q)}|^2 \}, \qquad \mathcal{F}^{(p,q)}(\alpha)  =  \{k : |\lambda_k^{(p,q)}|^2\stackrel{\cdot}{<} SNR^{-\alpha}\},\\
\mathcal{M}(\alpha) & =  \{ \mathbf{h} :  |h_{i}^{(p,q)}|^2 \stackrel{\cdot}{\leq}SNR^{-\alpha},  \forall i \in \{0,\ldots,\nu\} , \forall p \in \{1,\ldots,M_r\}, q \in \{1,\ldots,M_t\}  \}.
\end{align}
We have the following relations on the cardinality of these sets:
\renewcommand\theenumi{(\alph{enumi})}
\begin{enumerate}
\item Letting $\overline{\mathcal{G}^{(p,q)}}$ represent the
complement of the set $\mathcal{G}^{(p,q)}$, we have
\begin{equation}
  |\overline{\mathcal{G}^{(p,q)}}| \leq \nu \quad \forall p,q.
\end{equation}
In other words, at least $T_s-\nu$ of the $T_s$ taps in the frequency
domain, for each $(p,q)$, are (asymptotically) of magnitude
$\max\left(|h_{0}^{(p,q)}|^2,|h_{1}^{(p,q)}|^2,\ldots,|h_{\nu}^{(p,q)}|^2\right)$.
\item Given that $\mathbf{H} \in \overline{\mathcal{M}(\alpha)}$, for MISO
channel
\begin{equation}
\exists q \in \{1,2,\ldots,M_t\} \ s.\ t.\   |\mathcal{F}^{(1,q)}(\alpha)|\leq \nu
\end{equation}
and for a SIMO channel
\begin{equation}
\exists p \in \{1,2,\ldots,M_r\} \ s.\ t.\   |\mathcal{F}^{(p,1)}(\alpha)|\leq \nu
\end{equation}
\end{enumerate}
\end{lemma}

\begin{IEEEproof}
The tap coefficients in the frequency domain are given by,
\begin{align}
\lambda_{k}^{(p,q)} & =    \sum_{m=0}^{\nu} h_m^{(p,q)} e^{-\frac{2 \pi j}{T_s}k m} \qquad k =\{0,\ldots,(T_s-1)\}
\nonumber
\end{align}
Defining $\theta=e^{-\frac{2 \pi j}{T_s}}$ the above equation can be rewritten as,
\begin{align}
\lambda_{k}^{(p,q)} & =    \sum_{m=0}^{\nu} h_m^{(p,q)} \theta ^{k m} \qquad k =\{0,\ldots,(T_s-1)\} \\ \nonumber
& =  \left[\begin{array}{cccc} 1 & \theta^{k} & \ldots & \theta^{k\nu} \end{array} \right] 
%\left[\begin{array}{c} h_0^{(p,q)} \\ h_1^{(p,q)} \\ \ldots \\ h_\nu^{(p,q)} \end{array}\right]
\left[\begin{array}{cccc} h_0^{(p,q)} & h_1^{(p,q)} & \ldots & h_\nu^{(p,q)} \end{array}\right]^t
\nonumber
\end{align}
Take any set of $(\nu+1)$ coefficients in the frequency domain and
index this set by $\mathcal{K}=\{k_0,\ldots,k_{\nu} \}$ and define,
\begin{align}
\label{eqn:use_inverse}
\mathbf{\Breve{\Lambda}}^{(p,q)} & =  \left[\begin{array}{c} \lambda_{k_0}^{(p,q)} \\\lambda_{k_1}^{(p,q)} \\ \vdots \\ \lambda_{k_{\nu}}^{(p,q)} \end{array}\right] 
= \underbrace{\left[ 
\begin{array}{cccc} 
1 & \theta^{k_0} & \ldots & \theta^{k_0\nu} \\
1 & \theta^{k_1} & \ldots & \theta^{k_1\nu} \\
\vdots & & \ldots & \vdots \\
1 & \theta^{k_{\nu}} & \ldots & \theta^{k_{\nu}\nu} \\
\end{array}
\right] }_{\mathbf{V}} 
\underbrace{
\left[\begin{array}{c} h_0^{(p,q)} \\ h_1^{(p,q)} \\ \vdots \\ h_\nu^{(p,q)} \end{array}\right]
}_{\mathbf{h}} 
\end{align}
where $\mathbf{V}\in\mathbb{C}^{(\nu+1)\times(\nu+1)}$ is a full rank
Vandermonde matrix. Therefore, the inverse of $\mathbf{V}$ exists and
we denote it by $\mathbf{V}^{-1}=\mathbf{A}$ and let $a_{li}$
represent the element in the $l^{th}$ row and $i^{th}$ column. From
(\ref{eqn:use_inverse}) we have,
\begin{align*}
%\left[\begin{array}{c} h_0^{(p,q)} \\ h_1^{(p,q)} \\ \vdots \\ h_\nu^{(p,q)} \end{array}\right]
\left[\begin{array}{cccc} h_0^{(p,q)} & h_1^{(p,q)} & \ldots & h_\nu^{(p,q)} \end{array}\right]^t
=  \mathbf{V}^{-1} \Breve{\mathbf{\Lambda}} &,\,\,\,\, i.e. \,\,\,\,\, h_l^{(p,q)} =  \sum_{i=0}^{\nu} a_{li} \lambda_{k_i}^{(p,q)} \quad l=\{0,1,\ldots,\nu\}.
\end{align*}
Using the Cauchy-Schwartz inequality\footnote{$|\mathbf{u}^*
  \mathbf{v}|\leq \|u \|. \|v\|$.}, we get,
\begin{align*}
|h_l^{(p,q)}|^2 & = |\sum_{i=0}^{\nu} a_{li} \lambda_{k_i}^{(p,q)}|^2 \leq (\sum_{i=0}^{\nu} |a_{li} |^2)  (\sum_{i=0}^{\nu} |\lambda_{k_i}^{(p,q)} |^2) 
\end{align*}
Using the fact that $T_s$ is finite and does not grow with $SNR$ it
follows that the $\{a_{li}\}$ do not depend on $SNR$. Therefore, the
above inequality can be asymptotically written as
\begin{align}
\label{cauchy}
|h_l^{(p,q)}|^2  & \stackrel{\cdot}{\leq}   |\lambda_{k_0}^{(p,q)}|^2 + |\lambda_{k_1}^{(p,q)}|^2 + \ldots + |\lambda_{k_\nu}^{(p,q)}|^2.
\end{align}
Note that the above inequality holds for all $h_l^{(p,q)}$,
$l=0,\ldots,\nu$. Therefore, we get that for any set of $(\nu+1)$
coefficients in the frequency domain indexed by $\{k_0,\ldots,k_{\nu}
\}$,
\begin{align}
\label{cauchy1}
\max_{l\in\{ 0,1,\ldots,\nu\} }|h_l^{(p,q)}|^2  & \stackrel{\cdot}{\leq}   |\lambda_{k_0}^{(p,q)}|^2 + |\lambda_{k_1}^{(p,q)}|^2 + \ldots + |\lambda_{k_\nu}^{(p,q)}|^2  
\end{align}
From the Cauchy-Schwartz inequality note that,
\begin{align}
|\lambda_{k_0}^{(p,q)}|^2 + |\lambda_{k_1}^{(p,q)}|^2 + \ldots + |\lambda_{k_\nu}^{(p,q)}|^2  & = |\sum_{m=0}^{\nu} h_m^{(p,q)} \theta ^{k_0 m}|^2 +\ldots + |\sum_{m=0}^{\nu} h_m^{(p,q)} \theta ^{k_\nu m}|^2  \nonumber \\
\leq &  \left(\sum_{m=0}^{\nu} |h_m^{(p,q)}|^2 \right)\left(\sum_{m=0}^{\nu}|\theta ^{k_0 m}|^2 + \ldots +  \sum_{m=0}^{\nu}|\theta ^{k_{\nu} m}|^2\right)\nonumber \\
%= & (\nu+1)^2 (|h_0|^2+|h_1|^2+\ldots+|h_{\nu}|^2 )  \nonumber \\
%\doteq &  (|h_0^{(p,q)}|^2+|h_1^{(p,q)}|^2+\ldots+|h_{\nu}^{(p,q)}|^2 ) \nonumber \\
%\doteq &  \max_{l\in\{ 0,1,\ldots,\nu\} }|h_l^{(p,q)}|^2 \label{cauchy2}
\doteq &  (|h_0^{(p,q)}|^2+|h_1^{(p,q)}|^2+\ldots+|h_{\nu}^{(p,q)}|^2 ) \doteq  \max_{l\in\{ 0,1,\ldots,\nu\} }|h_l^{(p,q)}|^2 \label{cauchy2}
\end{align}
Combining equations (\ref{cauchy1}) and (\ref{cauchy2}) we get,
\begin{align}
\label{cauchy3}
 |\lambda_{k_0}^{(p,q)}|^2 + |\lambda_{k_1}^{(p,q)}|^2 + \ldots + |\lambda_{k_\nu}^{(p,q)}|^2 & \doteq  \max_{l\in\{ 0,1,\ldots,\nu\} }|h_l^{(p,q)}|^2 .
\end{align}
\begin{itemize}
\item We know from (\ref{cauchy2}) that for all k , $ |\lambda_k|^2 \stackrel{\cdot}{\leq} \max_{l\in\{ 0,1,\ldots,\nu\} }|h_l|^2 $. Since, $ \mathcal{G}^{(p,q)} =\{i : |\lambda_i^{(p,q)}|^2 \doteq \max_{l\in\{
0,1,\ldots,\nu\} }|h_l^{(p,q)}|^2\}$,
\begin{align*}
 |\lambda_k^{(p,q)}|^2 & \stackrel{\cdot}{<} \max_{l\in\{ 0,1,\ldots,\nu\}  }|h_l^{(p,q)}|^2 \qquad \forall k \in \overline{\mathcal{G}^{(p,q)}}.
\end{align*}
If $|\overline{\mathcal{G}^{(p,q)}}|>\nu $ then there exists a set
$\mathcal{K}=\overline{\mathcal{G}^{(p,q)}}$ of size at least $\nu+1$
such that,
\begin{align*}
 |\lambda_k^{(p,q)}|^2 & \stackrel{\cdot}{<}  \max_{l\in\{ 0,1,\ldots,\nu\} }|h_l^{(p,q)}|^2 \qquad  \forall k \in \mathcal{K}
\end{align*}
But this is a contradiction to equation (\ref{cauchy3}) and
therefore we have $|\overline{\mathcal{G}^{(p,q)}}|\leq \nu$ proving $(a)$.
\item For a MISO channel, given that $\mathbf{H} \in
  \overline{\mathcal{M}(\alpha)}$, we know that there exists at least one
  $(\hat{i},\hat{q})$ pair such that $|h_{\hat{i}}^{(1,\hat{q})}|^2
  \stackrel{\cdot}{>}\frac{1}{SNR^{\alpha}}$. For this
  $\hat{q}$, since
  $|\overline{\mathcal{G}^{(1,\hat{q})}}|\leq \nu$ it follows that
  $|\mathcal{F}^{(1,\hat{q})}(\alpha)|\leq \nu$.
\end{itemize}
\end{IEEEproof}

\section{ISI channels with single transmit antenna}
\label{chap5:sec:SIMTra}
Using results from Lemma \ref{lem:TapMag} we will show that uncoded QAM
transmission can be used to derive an alternative characterization of
the DM trade-off for the ISI channel. We will use uncoded QAM constellation
for transmission such that the minimum distance between any two points
in the constellation $d_{min}$ is such that $d_{min}^2 \stackrel{\cdot}{\geq} SNR^{(1-r)}$.
\par 
Consider a transmission scheme where the uncoded QAM symbols are
transmitted for a period $T_s-\nu$ followed by a padding with $\nu$
zeros. Since from the Lemma \ref{lem:TapMag} we know that
$|\mathcal{F}^{(1,1)}(r)|\leq \nu$, we ignore these $\nu$ channels and examine
the remaining $T_s-\nu$ channels in $\overline{\mathcal{F}^{(1,1)}(r)}$. We
can show that the distance between codewords in these channels is
still asymptotically larger than $SNR^{(1-r)}$. As the pairwise error
probability is a $Q$ function, we can show that the error probability
decays exponentially in SNR. This is summarized in the following
lemma, the proof of which is in the appendix.
\begin{lemma}
\label{lem:ExpDecOne}
Assume transmission from an uncoded QAM transmission ($\mathcal{X}$)
such that the minimum distance $d_{min}$ between any two points in the
constellation is lower bounded by $d_{min}^2\stackrel{\cdot}{\geq} SNR^{(1-r)}$. At each time instant one symbol is independently transmitted from the constellation for $T_s-\nu$ time instants followed by a padding with
$\nu$ zero symbols. For a finite period of communication (finite
$T_s$), given that $\mathbf{h} \in \overline{\mathcal{M}^(1-r)}$, the
error probability $P_e$ decays exponentially in $SNR$.
\end{lemma}

Note that Lemma \ref{lem:TapMag} and Lemma \ref{lem:ExpDecOne} can be combined
together to give an alternative proof of the diversity multiplexing
trade-off of the SISO ISI channel.
% as,
%\begin{align*}
%P_e(SNR) & =   P(\mathcal{M}(1-r)) P_{e}(SNR \mid  \mathbf{h} \in \mathcal{M}(1-r) ) +P(\overline{\mathcal{M}(1-r)}) P_{e}(SNR\mid %\mathbf{h} \in \overline{\mathcal{M}(1-r)}) \\
%& \stackrel{\cdot}{\leq}   P(\mathcal{M}(1-r)) + \left( 1-SNR^{-\nu(1-r)} \right)  P_{e}(SNR\mid \mathbf{h} \in  \overline{\mathcal{M}(1-%r)}) \\
%& \doteq   SNR^{-\nu(1-r)} \qquad \mbox{As}\ P(\mathcal{M}(1-r))\doteq SNR^{-\nu(1-r)},
%\end{align*}
%where the last equality is true from lemma \ref{lem:ExpDecOne}. Note that the
%effective rate of transmission because of the zero padding is
%$\frac{r(T_s-\nu)}{T_s}$ and to communicate at an effective rate of
%$r$ we need to send symbols from a QAM constellation of size
%$SNR^{\tilde{r}}$ where $\tilde{r}=\frac{rT_s}{T_s-\nu}$.

To prove the successive refinement of the SISO ISI trade-off we will
first prove a lemma analogous to Lemma \ref{lem:ExpDecOne} for superposition
coding, {\em i.e.}  the symbol transmitted at the $k^{th}$ instant is
the superposition of a symbol from $\mathcal{X}_H$, $\mathcal{X}_L$
given by,
\begin{align*}
x[k] & =  x_H[k]+x_L[k] \qquad {\rm where} \, \, x_H[k] \in \mathcal{X}_H,  \,\, x_L[k] \in \mathcal{X}_L
\end{align*}
\begin{lemma}
\label{lem:ExpDecTwo}
For $r_H,r_L\in[0,\frac{T_s-\nu}{T_s}]$ denote $\tilde{r}_H = r_H \frac{T_s}{T_s-\nu}$ and 
$\tilde{r}_L = r_L \frac{T_s}{T_s-\nu}$. Let $\mathcal{X}_H$ and $\mathcal{X}_L$ be QAM constellations of size
$SNR^{\tilde{r}_H}$ and $SNR^{\tilde{r}_L}$ with power constraint
$SNR$ and $SNR^{1-\beta}$ respectively, where $\beta>\tilde{r}_H$.
Assume uncoded superposition transmission such that at each time
instant symbols are independently chosen and superposed from each
constellation ($\mathcal{X}_H$, $\mathcal{X}_L$) for $(T_s-\nu)$ time
instants followed by a padding with $\nu$ zero symbols. For a finite
period of communication (finite $T_s$) given that $\mathbf{h} \in
\overline{\mathcal{M}(1-r_H)}$, the error probability of detecting the
set of symbols sent from the higher constellation, $\mathcal{X}_H$,
denoted by $P_e^{H}(SNR)$, decays exponentially in $SNR$.
\end{lemma}
The details of the proof are in the appendix. In this lemma we
critically use the fact that {\em all} except at most $\nu$ taps in
the frequency domain, are asymptotically of {\em equal} magnitude
($\max_{l\in\{ 0,1,\ldots,\nu\} }|h_l|^2$). Using these lemmas we will
prove the following theorem on the successive refinement of the SISO
ISI trade-off.
\begin{theorem}
\label{thm:SISORefi}
The diversity multiplexing trade-off for a $\nu$ tap point to point
SISO ISI channel is successively refinable, {\em i.e.}, for any
multiplexing gains $r_H$ and $r_L$ such that $r_H+r_L \leq
\frac{T_s-\nu}{T_s}$ the achievable diversity orders given by $D_{H}(r_H)$
and $D_{L}(r_L)$ are bounded as,
\begin{align}
(\nu+1)\left(1-\frac{T_s}{(T_s-\nu)}r_H\right)  & \leq  D_{H}(r_H) \leq  (\nu+1)\left(1-{r_H}\right) \label{higherlayer}, \\
(\nu+1)\left(1-\frac{T_s}{(T_s-\nu)}(r_{H}+r_L)\right) & \leq   D_{L}(r_{L}) \leq  (\nu+1)\left(1-(r_H+r_L)\right)  \label{higherlower}
\end{align}
where $T_s$ is finite and does not grow with SNR. 
\end{theorem}
\begin{IEEEproof}
To show the successive refinement we use superposition coding and
assume two streams with uncoded QAM codebooks for each stream, as in
\cite{DigTse05}. Assume that given a total power constraint $P$ we
allocate powers $P_{H}$ and $P_{L}$ to the high and low priority
streams respectively. We design the power allocation such that at high
signal to noise ratio we have $SNR_{H}\doteq SNR$ and $SNR_{L} \doteq
SNR^{1-\beta}$ for $\beta \in [0,1]$. Let $\mathcal{X}_H$ be QAM
constellation instant of size $SNR^{\tilde{r}_H}$ with minimum
distance $(d_{min}^H)^2=SNR^{1-\tilde{r}_H}$. Similarly let
$\mathcal{X}_L$ be a QAM constellation of size $SNR^{\tilde{r}_L}$
with minimum distance $(d_{min}^L)^2=SNR^{1-\beta-\tilde{r}_L}$, where
$\beta>\tilde{r}_H$. The symbol transmitted at the $k^{th}$ instant is
the superposition of a symbol from $\mathcal{X}_H$, $\mathcal{X}_L$
as in Lemma \ref{lem:ExpDecTwo}. It can be shown \cite{DigTse05} that even with the above superposition
coding, if $\beta>\tilde{r}_H$ the order of magnitude of the effective
minimum distance between two points in the constellation
$\mathcal{X}_H$ is preserved.

The upper bound in both (\ref{higherlayer}) and (\ref{higherlower}) is
trivial and follows from the matched filter bound. We will investigate
the lower bound in (\ref{higherlayer}). Superpose symbols from the
higher and lower layers for $(T_s-\nu)$ time instants and pad them
with $\nu$ zero symbols at the end.  With this particular transmission
scheme, given that $\mathbf{h} \in \overline{\mathcal{M}(1-r_H)}$, we
know from Lemma \ref{lem:ExpDecTwo} that the error probability decays
exponentially in $SNR$. Therefore,
\begin{align}
P_{e}^H(SNR) & =  P(\mathcal{M}(1-r_H)) P_{e}(SNR \mid \mathbf{h} \in \mathcal{M}(1-r_H) ) + P(\overline{\mathcal{M}(1-r_H) }) P_{e}(SNR\mid \mathbf{h} \in \overline{ \mathcal{M}(1-r_H) } ) \nonumber\\
& \leq   P(\mathcal{M}(1-r_H))  + P(\overline{\mathcal{M}(1-r_H) }) P_{e}(SNR\mid \mathbf{h} \in \overline{ \mathcal{M}(1-r_H) }  )  \nonumber \\
&  \doteq  SNR^{-(\nu+1)(1-\tilde{r}_H)} + \left( 1-SNR^{-(\nu+1)(1-\tilde{r}_H)} \right)   P_{e}(SNR\mid \mathbf{h} \in \overline{ \mathcal{M}(1-r_H) }  ). \label{eq:ErrProTer2}
\end{align}
For decoding the higher layer we treat the signal on the lower layer
as noise. Given that $\mathbf{h} \in \overline{\mathcal{M}(1-r_H)}$ and
choosing $\beta>\tilde{r}_H$ we conclude from Lemma
\ref{lem:ExpDecTwo} that the second term in (\ref{eq:ErrProTer2})
decays exponentially in $SNR$. Therefore,
\begin{align}
P_{e}^H(SNR) & \stackrel{\cdot}{\leq} SNR^{-(\nu+1)(1-\tilde{r}_H)}
\end{align}
or equivalently, $(\nu+1)\left(1-\frac{T_s}{(T_s-\nu)}r_H\right)  \leq  D_{H}(r_H)$ for 
communication at a rate of $r_H =\frac{(T_s-\nu)}{T_s}\tilde{r}_H$. Once we have decoded the upper layer we subtract its contribution from
the lower layer. Note that the minimum distance between two points in the QAM
constellation for the lower layer is given by $(d_{min}^L)^2 =SNR^{1-\beta}SNR^{-\tilde{r}_L}=SNR^{1-\beta-\tilde{r}_L}$. 
 Using the set $\mathcal{M}(1-\tilde{r}_L-\beta)$, Lemma \ref{lem:ExpDecOne} and by taking $\beta$ arbitrarily
close to $\tilde{r}_H$,  we can conclude (\ref{higherlower}). Comparing this with Theorem \ref{thm:ISIDivMulTraRep} we can see that
the diversity multiplexing trade-off for the SISO ISI channel is
successively refinable.
\end{IEEEproof}
The intuition that was used in deriving the successive refinement of
the SISO trade-off for ISI channels was that given that
$\mathbf{h}\in\mathcal{M}(1-r)$ at most $\nu$ taps in the frequency
domain are zero and the remaining are ``good'' and of the same
magnitude. This intuition can also be carried over to show the
successive refinability of the SIMO channel with $M_r$ receive
antennas and one transmit antenna as well.

\section{Successive refinement of MISO ISI channel}
\label{chap5:sec:MISTra}
\begin{figure}
\centering
\includegraphics[scale=0.8]{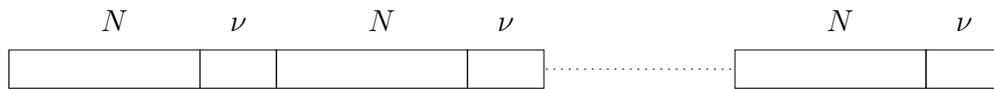}
\caption{Coding Scheme for MISO channels}
\label{fig:ComSch}
\end{figure}

In Section \ref{chap5:sec:SIMTra} we saw that superposition of uncoded
QAM constellations followed by zero padding was sufficient to prove
the successive refinability of the SISO/SIMO ISI trade-off. The proof
of successive refinability of the MISO channel requires a more
sophisticated coding strategy as discussed in this section \cite{DusDig08b}. We will
consider the same model as in Section \ref{chap5:sec:StrObs} and
consider a slight variant of the scheme where we transmit for a period
of $(T_s-\nu)$ followed by padding with $\nu$ zeros. The variant is
that instead of doing this once we repeat the process of transmission
for $(T_s-\nu)$ followed by $\nu$ zeros for $T_b=T_s M_t$ symbols for
a total communication period of $T=T_b T_s$. The scheme is as shown in
the Figure \ref{fig:ComSch}.

We rearrange the received $T$ symbols in a matrix form
$\Ybf\in\mathbb{C}^{T_s \times T_b}$ where,
\begin{align}
\Ybf & = \left[ \begin{array}{cccc} \mathbf{y}^{(1)}_{[0,T_s-1]} & \mathbf{y}^{(1)}_{[T_s,2T_s-1]}& \ldots & \mathbf{y}^{(1)}_{[(T_b-1)T_s,T_bT_s-1]} \end{array} \right].
\end{align}
Denote
$\bfbreveX^{(i)}\in\mathbb{C}^{T_s\times T_b}$ to be the symbols
transmitted by the $i^{th}$ transmit antenna over the period of
communication {\em i.e.},
\begin{align*}
\bfbreveX^{(i)} 
& = 
\left[\begin{array}{cccc} 
\xbf^{(i)}_{[0,T_s-\nu-1]} &  \mathbf{x}^{(1)}_{[T_s,2T_s-\nu-1]} & \ldots & \mathbf{x}^{(i)}_{[(T_b-1)T_s,T_bT_s-\nu-1]}\\  
\mathbf{0}_{\nu  \times 1}  & \mathbf{0}_{\nu  \times 1} & \ldots & \mathbf{0}_{ \nu \times 1} 
\end{array} \right]
 = 
\left[ \begin{array}{c}
{\Xbf^{(i)}} \\
\mathbf{0}_{\nu \times T_b}
\end{array} \right],
\end{align*}
for $ i\in \{ 1,2,\ldots,M_t \}.$ Similar to (\ref{eq:CircChan3}) we can rearrange the rows and columns
and the received symbols can be written as,
\begin{align}
\label{eq:MISO1}
\Ybf & =
    \left[
      \begin{array}{cccc}
        \mathbf{H}^{(1,1)} & \mathbf{H}^{(1,1)} & \ldots & \mathbf{H}^{(1,M_t)} 
     \end{array}
      \right]
    \left[ \begin{array}{cccc}
        \bfbreveX^{(1)t} &
        \bfbreveX^{(2)t} &
        \ldots & 
        \bfbreveX^{(M_t)t}
      \end{array}
      \right]^t
    +\Zbf
%    \left[
%      \begin{array}{cccc}
  %      \mathbf{H}^{(1,1)} & \mathbf{H}^{(1,1)} & \ldots & \mathbf{H}^{(1,M_t)} 
%     \end{array}
 %     \right]
 %   \left[ \begin{array}{c}
  %      \bfbreveX^{(1)} \\
  %      \bfbreveX^{(2)} \\
   %     \vdots \\ 
  %      \bfbreveX^{(M_t)}
  %    \end{array}
  %    \right]
  %  +\Zbf
\end{align}
where $\mathbf{H}^{(1,1)},\ldots,\mathbf{H}^{(1,M_t)} \in
\mathbb{C}^{T_s \times T_s }$ are circulant matrices given in
(\ref{eq:DefCircH}). Decomposing them in frequency domain notation, as
in (\ref{eq:DefLambda2}), and premultiplying $\mathbf{Y}$ by
$\mathbf{Q}^*$ we can rewrite (\ref{eq:MISO1}) as
\begin{align}
\bftildeY & =\Qbf^* \Ybf 
 = 
\Qbf^* \left[  
\begin{array}{cccc}
\Qbf\bfLambda^{(1,1)} \Qbf^{*} & \Qbf \bfLambda^{(1,2)} \Qbf^{*} & \ldots & \Qbf\bfLambda^{(1,M_t)} \Qbf^{*} 
\end{array}
\right]
\left[ \begin{array}{c}
\bfbreveX^{(1)}\\
\bfbreveX^{(2)}\\
\vdots\\
\bfbreveX^{(M_t)}\\
\end{array}
\right]
+\Qbf^* \Zbf \nonumber 
\\
& = 
\left[
\begin{array}{cccc}
\bfLambda^{(1,1)} & \bfLambda^{(1,2)} & \ldots & \bfLambda^{(1,M_t)} 
\end{array}
\right]
\left[ \begin{array}{c}
\bftildeQ^{*} \Xbf^{(1)}  \\ 
\bftildeQ^{*} \Xbf^{(2)}  \\ 
\vdots \\
\bftildeQ^{*} \Xbf^{(M_t)}  
\end{array}
\right]
+\Qbf^* \Zbf \nonumber  \\
& = \underbrace{\left[
\begin{array}{cccc}
\bfLambda^{(1,1)} & \bfLambda^{(1,2)} & \ldots & \bfLambda^{(1,M_t)} 
\end{array}
\right]
}_\bfLambda 
\left[\begin{array}{cccc}
\bftildeQ^{*} & 0 & \ldots &0 \\
0 & \bftildeQ^{*}  & \ldots & 0\\
\vdots & & \ddots & \\
0 & \ldots & 0 & \bftildeQ^{*}
\end{array}
\right]
\underbrace{
\left[ \begin{array}{c}
\Xbf^{(1)}  \\ 
\Xbf^{(2)}  \\ 
\vdots \\
\Xbf^{(M_t)}  
\end{array}
\right]
}_\mathbf{X}
+\underbrace{ \Qbf^* \Zbf }_\bftilde{Z} \label{defn_X}
\end{align}
where $\bftildeQ^{*}\in \mathbb{C}^{T_s\times (T_s-\nu)} $ is a matrix
obtained by deleting the last $\nu$ columns and $\bftilde{Z}$ still
has i.i.d. Gaussian entries. Observe that since $\bftildeQ^{*}$ is a
$T_s \times (T_s-\nu) $ Vandermonde matrix, it is a full rank matrix. 

Let $\kappa\in \{1,\ldots,M_t\}$ represent the antenna which has the
maximum tap coefficient out of all the $M_T(\nu+1)$ coefficients in
the time domain, {\em i.e.},
\begin{align*}
  \max_{p \in \{1,\ldots,M_t\},l\in\{0,\ldots,\nu\}} |h_l^{(1,p)}|^2 & \stackrel{\cdot}{\leq} max\left(|h_0^{(1,\kappa)}|^2,\ldots,|h_{\nu}^{(1,\kappa)}|^2\right).
\end{align*}
Define a selection matrix $\Sbf \in \mathbb{C}^{(T_s-\nu) \times T_s}$
such that,
\begin{align*}
\lefteqn{
\Sbf \bfLambda 
\left[\begin{array}{cccc}
\bftildeQ^{*} & 0 & \ldots & 0 \\
0 & \bftildeQ^{*} & \ldots & 0\\
\vdots & & \ddots & \\
0 & \ldots & 0 & \bftildeQ^{*}
\end{array}
\right] 
%& = \left[
 = \left[
\begin{array}{cccc}
\Sbf \bfLambda^{(1,1)} \bftildeQ^{*}  & \Sbf \bfLambda^{(1,2)}\bftildeQ^{*}  & \ldots & \Sbf \bfLambda^{(1,M_t)}\bftildeQ^{*} 
\end{array}
\right]
} &
\\
& \stackrel{(a)}{=}
\left[
\begin{array}{cccc}
\bfhatLambda^{(1,1)} \bfhat{Q}^*  & \bfhatLambda^{(1,2)} \bfhat{Q}^* & \ldots & \bfhatLambda^{(1,M_t)} \bfhat{Q}^* 
\end{array}
\right]
%& 
%\\
%& = 
=
\underbrace{
\left[
\begin{array}{cccc}
\bfhatLambda^{(1,1)} & \bfhatLambda^{(1,2)} & \ldots & \bfhatLambda^{(1,M_t)} 
\end{array}
\right]
}_{\bfhatLambda}
\underbrace{
\left[\begin{array}{cccc}
\bfhat{Q}^* & 0 & \ldots &0 \\
0 & \bfhat{Q}^* & \ldots & 0\\
\vdots & & \ddots & \\
0 & \ldots & 0 & \bfhat{Q}^*
\end{array}
\right]
}_
{\mathbf{\breve{Q}^*}}
\end{align*}
where, $\bfhatLambda^{(1,i)}\in\mathbb{C}^{(T_s-\nu) \times
  (T_s-\nu)}$ and in particular,
$\bfhatLambda^{(1,\kappa)}=diag\left( \{\lambda_l^{(1,\kappa)}:l\in \mathcal{G}^{(1,\kappa)}\}\right). $
The step $(a)$ is valid above as $\bfhatLambda^{(1,i)}$ is a diagonal
matrix. Therefore $\mathbf{S} \bfhatLambda^{(1,i)}$ will have exactly
$T_s-\nu$ columns with non zero entries and will have $\nu$ columns
with all zero entries. Therefore $\Sbf \bfhatLambda^{(i)}
\bftildeQ^{*}$ can be written as $\bfhatLambda^{(i)} \bfhat{Q}^*$
where $\bfhatLambda^{(i)}$ is as defined above. Also $\bfhat{Q}^*\in
\mathbb{C}^{(T_s-\nu) \times (T_s-\nu)}$ is the matrix $\bftildeQ^{*}$
with the $\nu$ rows corresponding to $\{\lambda_l^{(1,\kappa)}:l\in
\overline{ \mathcal{G}^{(1,\kappa)}}\}$ deleted and
$\mathbf{\breve{Q}^*}\in \mathbb{C}^{(T_s-\nu)M_t \times (T_s-\nu)M_t}
$. Using the same selection matrix for the whole block of $T_b$
symbols we have,
\begin{align}
\bfhat{Y} & =  \mathbf{S} \bftildeY  = \bfhatLambda \mathbf{\breve{Q}^*} \mathbf{X}+ \bfhat{Z}, \label{eq:AfterSele}
\end{align}
where $\bfhat{Z} $ is still iid Gaussian as it is obtained by deleting
$\nu$ rows from $\bftilde{Z}$. Now we will impose constraints on the
codewords $\mathbf{X}$ to ensure the diversity embedding for MISO ISI
channels.

\subsection{Codebook Constraints}
\label{chap5:sec:subsec:CodCon}
For transmission we consider a superposition of two $(T_s-\nu)M_t
\times (T_s-\nu)M_t$ codebooks $\mathcal{X}_H$ and $\mathcal{X}_L$ of
rates $ (T_s-\nu)r_H$ and $(T_s-\nu)r_L$ respectively satisfying the
following design criteria:
\begin{enumerate}
\item For $r_H\in[0,1]$ and defining $\Delta \Xbf_H=
\Xbf_H-\Xbf_H^{'}\neq 0$, for all $\Xbf_H,\Xbf_H^{'}\in\mathcal{X}_H $
we require that,
\begin{align}
\| \Xbf_H \|_F^2 \leq (T_s-\nu)M_t \, SNR & \leq (T_s-\nu)M_t  T_s \,SNR    \label{constraint2} \\
\min_{  \Delta \Xbf_H }  det\left(\Delta \Xbf_H \Delta \Xbf_H^{*}\right) & \stackrel{\cdot}{\geq} SNR^{(T_s-\nu)-(T_s-\nu)r_H} \label{constraint3}
\end{align}
\item For $r_L,\beta \in [0,1]$ and defining $\Delta \Xbf_L=
\Xbf_L-\Xbf_L^{'}\neq 0$, for all $\Xbf_L,\Xbf_L^{'}\in\mathcal{X}_L$
we require that,
\begin{align}
\| \Xbf_L \|_F^2  \leq (T_s-\nu)M_t\, SNR^{1-\beta} \leq  (T_s-\nu)M_t T_s \,SNR^{1-\beta}  & \label{constraint2_lower} \\
\min_{ \Delta \Xbf_L}  det\left(\Delta \Xbf_L \Delta \Xbf_L^{*}\right) \stackrel{\cdot}{\geq} SNR^{(T_s-\nu)M_t-(T_s-\nu)\beta-(T_s-\nu)r_L}& \label{constraint3_lower}
\end{align}
\end{enumerate}
We will use superposition coding from $\mathcal{X}_H,\mathcal{X}_L$ so
that $\mathbf{X}  = \Xbf_H+\Xbf_L.$ A particular set of codebooks satisfying these properties is
constructed in \cite{EliKum06} therefore establishing existence of
codes with these properties. Since we are padding every $T_s-\nu$
symbols with $\nu$ zeros, if the code $\mathcal{X}_H$ is designed with
rate $(T_s-\nu)r_H$ the effective rate of communication is
$\frac{(T_s-\nu)r_H}{T_s}$. Also, because of the energy constraint we
have,
\begin{align}
\|\Delta \Xbf_H \|_F^2  = tr \left(\Delta \Xbf_H \Delta \Xbf_H^{*}\right) \stackrel{\cdot}{\leq} SNR, & \,\,\,\,\qquad
\|\Delta \Xbf_L \|_F^2  = tr \left(\Delta \Xbf_L \Delta \Xbf_L^{*}\right) \stackrel{\cdot}{\leq} SNR^{1-\beta} \label{constraint4}
\end{align}

\subsection{Successive Refinement}
Assuming transmission using the superposition coding as in Section
\ref{chap5:sec:subsec:CodCon}, (\ref{eq:AfterSele}) is equivalent to,
\begin{align}
\bfhat{Y} & =  \bfhatLambda \mathbf{\breve{Q}^*} \mathbf{X}_H+ \bfhatLambda \mathbf{\breve{Q}^*} \mathbf{X}_L + \bfhat{Z} 
\end{align}
For decoding the higher layer we treat the signal on the lower layer
as noise. Representing $\|\cdot\|$ to be the Frobenius norm, the
decoding rule for $\Xbf_H$ is given by,
\begin{align}
\bfhat{X}_H & =  \argmin_{\Xbf_H} \| \bfhat{Y}-\bfhatLambda \bfbreveQ^{*} \Xbf_H \|^2. \label{decoding_rule}
\end{align}
Using this decoding rule, the pairwise error probability can be upper
bounded as in the following lemma.
\begin{lemma} 
\label{lem:Decoder}
The pairwise error probability of detecting the sequence $\Xbf_H^{'}$
given that $\Xbf_H$ was transmitted is upper bounded by,
%\begin{align*}
%P_e(\Xbf_H \rightarrow \Xbf_H^{'}|\bfh,\Xbf_L) & \leq Q \big( \| \bfhatLambda \bfbreveQ^{*} \Delta\Xbf_H\| -  2\sum_{i=1}^{NM_t} \| \bfhatLambda \bfbreveQ^{*}\bfx_L^{(i)} \|   \big) 
%\end{align*}
\begin{align}
P_e(\Xbf_H \rightarrow \Xbf_H^{'}|\hbf,\Xbf_L) & \leq Q \left( \| \bfhatLambda \bfbreveQ^{*} (\Xbf_H-\Xbf_H^{'})\| - 2\sum_{i=1}^{(T_s-\nu)M_t} \| \bfhatLambda \bfbreveQ^{*}\xbf_L^{(i)} \|   \right),  \label{pep2}
\end{align}
where $\xbf_L^{(i)}$ is the $i^{th}$ column of $\Xbf_L$.
\end{lemma}
The proof of this lemma can be done using standard techniques and the
details are in the appendix. Note that the error probability depends
on the Frobenius norm of $(\bfhatLambda \bfbreveQ^{*}
(\Xbf_H-\Xbf_H^{'}))$, which is related to the
singular values of $\bfbreveQ^{*} (\Xbf_H-\Xbf_H^{'})$ and
$\bfhatLambda$. Therefore, we get bounds on the singular values of
these two matrices in the Lemmas \ref{SVDQDeltaX} and \ref{SVDLambda} and defer
the proof to the appendix.
\begin{lemma}
\label{SVDQDeltaX}
Representing $\Delta \Xbf_H= \Xbf_H-\Xbf_H^{'}\neq 0$ for
$\Xbf_H,\Xbf_H^{'}\in\mathcal{X}_H$, $\bfbreveQ^* \Delta
\Xbf_H \Delta \Xbf_H^{*} \bfbreveQ$ can be written as,
\begin{align}
\bfbreveQ^* \Delta \Xbf_H \Delta \Xbf_H^{*} \bfbreveQ & = \mathbf{R} \mathbf{D}_2^2 \mathbf{R}^*, \label{SVD_second}
\end{align}
where $\mathbf{R}$ is a unitary matrix chosen such that  $\mathbf{D}_2^2  = diag\left(\xi_1^2,\xi_2^2,\ldots,\xi_{(T_s-\nu)M_t}^2 \right)$ and $\xi_1^2 \leq \xi_2^2 \leq \ldots \leq \xi_{(T_s-\nu)M_t}^2$. Then we have the following bounds on $\xi$,
\begin{align}
\displaystyle \prod_{i=1}^{T_b} \xi_i^2 & \stackrel{\cdot}{\geq} SNR^{(T_s-\nu)M_t-(T_s-\nu)r} \label{lower_bound_product} \\
\displaystyle \max_{i\in \{1,\ldots,(T_s-\nu)M_t\}} (\xi_i^2) & \stackrel{\cdot}{\leq} SNR. \label{upper_bound_max}
\end{align}
\end{lemma}
The decomposition of $\bfhat{\Lambda} \bfhat{\Lambda}^{*}$ can be used
to get the representation of $\bfhatLambda^* \bfhatLambda$ as
summarized in the following lemma.
\begin{lemma}
\label{SVDLambda}
$\bfhatLambda^* \bfhatLambda \in \mathbb{C}^{T_b \times T_b}$ can be
represented as $\bfhatLambda^* \bfhatLambda = \mathbf{V}^* \mathbf{D}_3^2 \mathbf{V}$,
where $\mathbf{V} \in \mathbb{C}^{(T_s-\nu)M_t \times (T_s-\nu)M_t}$
is a unitary matrix chosen such that $\mathbf{D}_3^2  = diag \left( \gamma_1^2, \gamma_2^2,\ldots,\gamma_{(T_s-\nu)}^2,\ldots,\gamma_{(T_s-\nu)M_t}^2 \right)$ and $\gamma_1^2 \geq \gamma_2^2 \ldots \geq \gamma_{(T_s-\nu)M_t}^2$. Then,
\begin{align}
\gamma_{i}^2 & \doteq \max_{ i\in\{0,1,\ldots,\nu\} } |h_i^{(1,\kappa)}|^2 =  \lambda^2  \qquad  i \leq (T_s-\nu)  \label{value_gamma}
\end{align}
and $\gamma_{i}^2=0$ for $i>(T_s-\nu)$. 
\end{lemma}
Combining these two lemmas and using the optimal decoder derived in
Lemma \ref{lem:Decoder} we can derive the following lemma on the
exponential decay of error probability:
\begin{lemma}
\label{lem:MISOEXPDecay}
Consider communication over a $\nu$ tap MISO ISI channel using
codewords from $\mathcal{X}_H$ and $\mathcal{X}_L$ as described in
section \ref{chap5:sec:subsec:CodCon}. For a finite period of
communication (finite $T_s T_b$), given that $\hbf \in
\overline{\mathcal{M}(1-r_H)}$, the error probability of detecting the
set of symbols sent from the higher constellation ($\mathcal{X}_H$)
denoted by $P_e^{H}(SNR)$ decays exponentially in $SNR$.
\end{lemma}
With these lemmas for the MISO channel, the successive refinement of
the MISO channel can be stated as:
\begin{theorem}
\label{thm:MISORefi}
The diversity multiplexing trade-off for a $\nu$ tap point to point
MISO ISI channel is successively refinable, {\em i.e.}, for any
multiplexing gains $r_H$ and $r_L$ such that $r_H+r_L \leq
\frac{T_s-\nu}{T_s}$ the achievable diversity orders given by $D_{H}(r_H)$
and $D_{L}(r_L)$ are bounded as,
\begin{align}
(\nu+1)M_t\left(1-\frac{T_s}{(T_s-\nu)}r_H\right)  & \leq  D_{H}(r_H) \leq  (\nu+1)M_t\left(1-{r_H}\right), \\
(\nu+1)M_t\left(1-\frac{T_s}{(T_s-\nu)}(r_{H}+r_L)\right) & \leq   D_{L}(r_{L}) \leq  (\nu+1)M_t\left(1-(r_H+r_L)\right)
\end{align}
where $T_s$ is finite and does not grow with SNR. 
\end{theorem}
The details of the proof are similar to the proof of Theorem
\ref{thm:SISORefi}.

\section{Discussion}
\label{sec:Disc}
The constraints on the codebook in Section
\ref{chap5:sec:subsec:CodCon} specializes to simpler cases for
particular channels. Thus, the inequalities in (\ref{constraint3}) and
(\ref{constraint3_lower}) are sufficient but not necessary conditions
for successive refinability. For example, the coding scheme to achieve
the successive refinement of the D-M trade-off in \cite{DigTse05} for
transmission over flat fading channel is a special case of the
codebook in Section \ref{chap5:sec:subsec:CodCon} with $\nu=0$ and
$T_s=1$. Similarly the uncoded QAM constellations used for the
SISO/SIMO ISI channel in Section \ref{chap5:sec:SIMTra} can be shown
to satisfy the constraints in equations (\ref{constraint3}) and
(\ref{constraint3_lower}) with $T_b=1$.

Theorems \ref{thm:MISORefi} and \ref{thm:SISORefi} implies that for
ISI fading channels with a single degree of freedom, it is possible to
design (asymptotically in SNR) ideal {\em opportunistic} codes.  The
existence of (almost) ideal opportunistic codes is surprising since
one would have expected the behavior for the ISI channel to be closer
to the flat-fading multiple-degrees-of-freedom case, where the D-M
trade-off was not successively refinable \cite{DigTse06}.

\begin{figure}%[htp]
     \centering \subfigure[Medium SNR]{
      \includegraphics[scale=0.5]{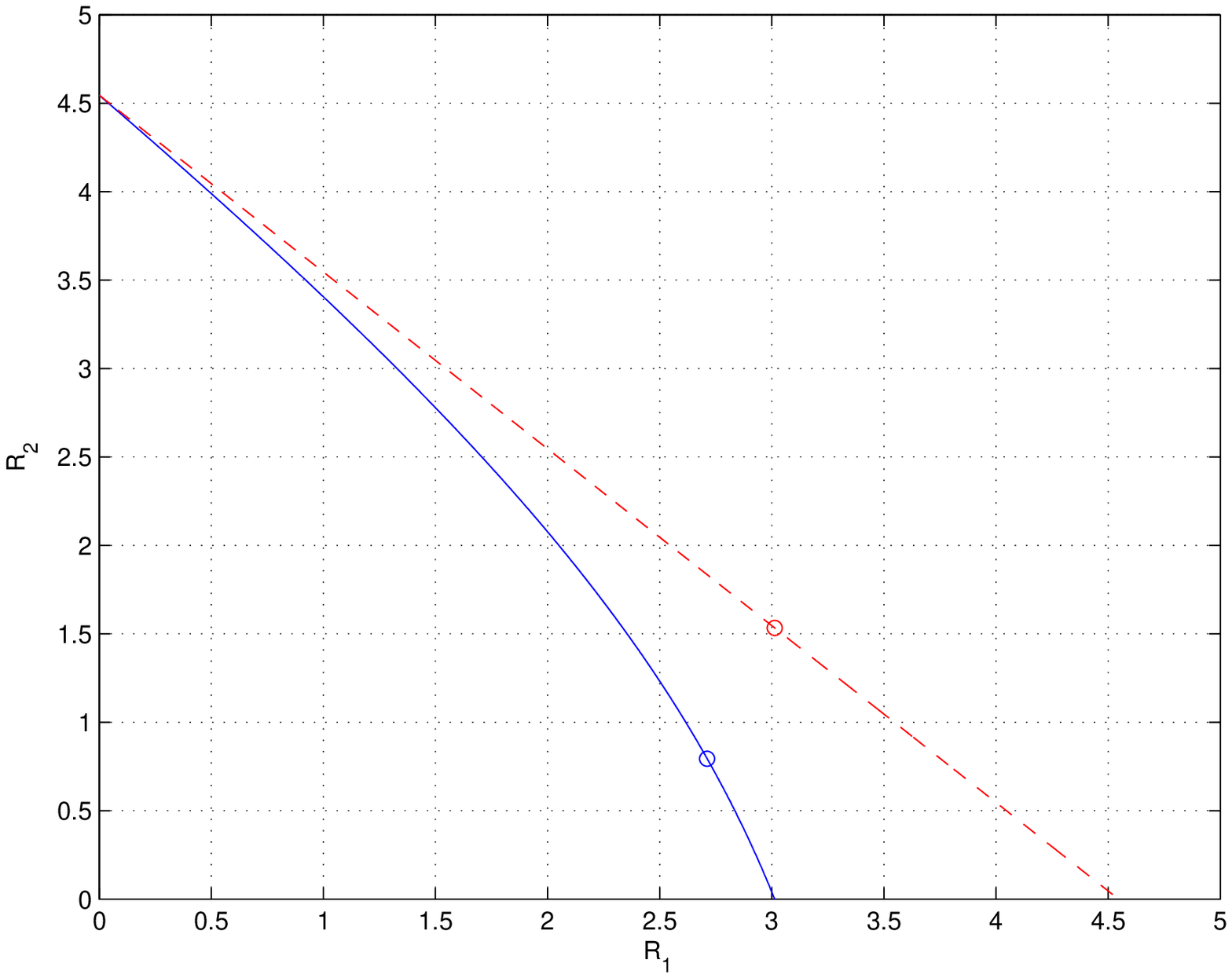}}
     \hspace{.3in}
     \subfigure[Higher SNR]{
              \includegraphics[scale=0.5]{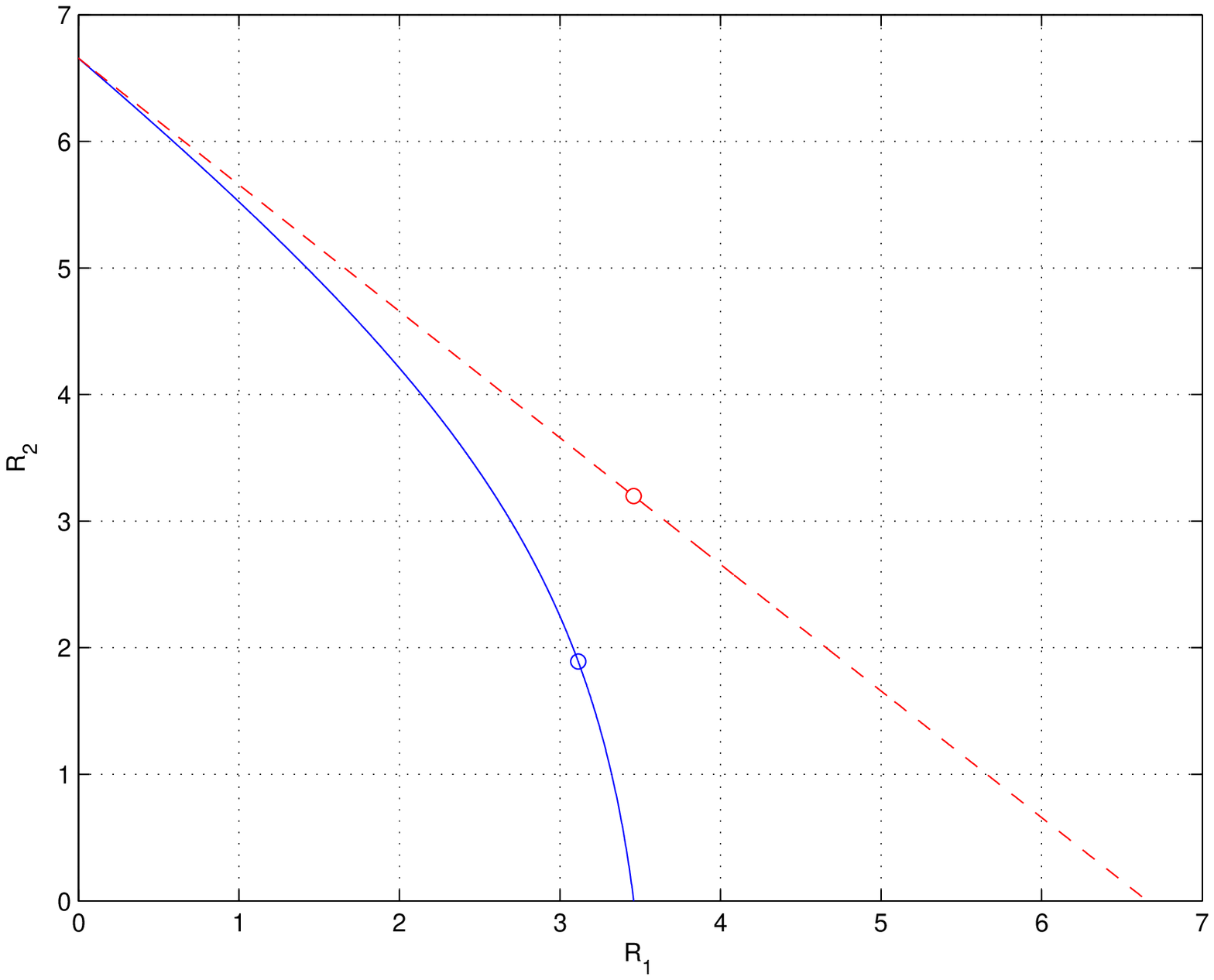}
}
\caption{The rate region illustrated for the scalar Gaussian broadcast channel with
the rate for the weaker channel on the x-axis and the stronger channel on the y-axis.
The rates are illustrated for $M_t=2,M_r=1$, and channels for given outage probabilities. 
The rates in Figure (a) are for 20 dB SNR, and typical channels corresponding to outage 
of $p_H=10^{-2},p_L=10^{-1}$. The rates in (b) are for a higher SNR of 30 dB, and we notice
that the region looks closer to a trapeziod ({\em i.e.,} the curve hugs the 45 degree
line shown for illustration, and departs almost vertically downwards). This shows that
for a small reduction in the rate for the worse channel, a large increase for the better
channel can be obtained. Asymptotically this trapezoidal shape gives the intuition for
the successive refinement of diversity property since the reduction needed for the worse
channel is small (in terms of multiplexing rate) and still attaining the optimal sum rate.}
\label{fig:Broadcast}
\end{figure}

We can also interpret the successive refinability by using the rate
region for the broadcast channel with user channels corresponding to
the typical error events of the corresponding diversity levels as
shown in Figure \ref{fig:Broadcast}. It demonstrates that as SNR grows
the shape of the Gaussian broadcast capacity region becomes closer to a
trapezoid. This implies that by reducing the rate slightly for the
high priority user (worse channel) we can significantly increase the
rate for the low priority user (better channel). The figure is plotted
for different SNR levels, and the result shows that asymptotically the
loss in rate for this exchange becomes very small. This gives an
engineering interpretation of the successive refinement result.

This paper demonstrated the successive refinement of diversity for ISI
fading channels with a single degree of freedom. However, many
questions remain open. Given the result of \cite{DigTse06} for flat
fading channels, it is natural to expect that the successive
refinement property will not hold for MIMO ISI fading
channels. However, there is an advantage of layering information, and
characterizing the rate-diversity tuples for MIMO channels would be an
important open question. Other issues including practical decoding
schemes and impact of this on multimedia applications would be natural
avenues of future enquiry.

%\subsection{Specialization of the Coding Scheme}

\begin{appendix}
\section{Proof of Lemmas for SISO ISI channel}
\subsection{Proof of Lemma \ref{lem:ExpDecOne}}
\begin{IEEEproof}
Denote the transmitted sequence of length $T_s-\nu$ by $\mathbf{x}\in
\mathcal{X}^{(T_s-\nu)}$, the $\nu$ zero symbols padded at the end by
$\mathbf{0}_{\nu \times 1}$. As a result of the zero padding,
proceeding along the lines of the proof of Lemma
\ref{thm:ISIDivMulTraRep} we can write the $T_s$ length received
vector as,
\begin{align}
\mathbf{y} & =  \mathbf{Q\Lambda Q^*} \left[\begin{array}{c} \mathbf{x} \\ \mathbf{0}_{\nu \times 1} \end{array}\right]+\mathbf{z} \label{use_for_simo}
\end{align}
where $\Qbf$ is a DFT matrix with the entries given by,
\begin{align}
\label{eq:DefFFT}
\Qbf_{p,q} & = e^{-\frac{2\pi j }{T_s} p q } \ {\rm for} \,\, 0 \leq p \leq T_s-1, \,\, 0\leq q \leq T_s-1
\end{align}
and $\mathbf{\Lambda}$ is a diagonal matrix with elements given by
\begin{align}
\label{eq:DefLambda}
\mathbf{\Lambda} & =  diag\left\{\lambda_{k} : \lambda_{k} =\sum_{m=0}^{\nu}h_m e^{-\frac{2 \pi j}{T_s}k m}  \right\}
\end{align}
for $ k =\{0,\ldots,(T_s-1)\}$. Note that $\mathbf{Q}$ is a
Vandermonde matrix which implies that it is a full rank
matrix. Multiplying the received vector by $\mathbf{Q^*}$ we get,
\begin{align*}
\mathbf{\tilde{y}} & =  \mathbf{Q^* y} = \mathbf{\Lambda} \mathbf{Q^*\left[\begin{array}{c} \mathbf{x} \\ \mathbf{0}_{\nu \times 1} \end{array}\right] } + \mathbf{ Q z} 
 = \mathbf{\Lambda} \mathbf{\tilde{Q}^*} \mathbf{x} + \mathbf{\tilde{z}} 
\end{align*}
where $\mathbf{\tilde{Q}^*}\in \mathbb{C}^{T_s\times (T_s-\nu)} $ is a
matrix obtained by deleting the last $\nu$ columns. Since
$\mathbf{\tilde{Q}^*}$ is also a Vandermonde matrix we conclude that
it has rank $(T_s-\nu)$.

From Lemma \ref{lem:TapMag}, given that
$\mathbf{h}\in\overline{\mathcal{M}(1-r)}$ we have that
$\mathcal{F}^{(1,1)}(r)\leq \nu$, {\em i.e.}, at most $\nu$ taps of
the available $T_s$ taps in the frequency domain can be of magnitude,
$|\lambda_{k}|^2\stackrel{\cdot}{\leq} SNR^{-(1-r)}. $ Define a selection matrix $\mathbf{S} \in \mathbb{C}^{(T_s-\nu) \times
T_s}$ such that,
\begin{align*}
  \mathbf{S}\mathbf{\Lambda} \mathbf{\tilde{Q}}^{*} &  =  \bfhat{\Lambda} \bfhat{Q}
\end{align*}
where, $\bfhat{\Lambda}\in\mathbb{C}^{(T_s-\nu) \times (T_s-\nu)}$ and,
$\bfhat{\Lambda} = diag\left( \{\lambda_l:l\in \overline{\mathcal{F}^{(1,1)}(r)}\}\right).$ 
Similarly $\bfhat{Q}\in \mathbb{C}^{(T_s-\nu) \times (T_s-\nu)}$ is
the matrix $\mathbf{\tilde{Q}}$ with the $\nu$ rows corresponding to
$\{\lambda_l:l\in \mathcal{F}^{(1,1)}(r)\}$ deleted. Note that $\bfhat{Q}$ is
still a full rank (rank $(T_s-\nu)$) Vandermonde matrix and denoting
the singular values of $\bfhat{\Lambda} \bfhat{Q}$ by $\gamma_k$ we
have, $\gamma_k \stackrel{\cdot}{>}SNR^{-(1-r)}.$ Using this selection matrix we have,
\begin{align}
\bfhat{y} & =  \mathbf{S} \mathbf{\tilde{y}}  
= \bfhat{\Lambda} \bfhat{Q}\mathbf{x} + \bfhat{z}. \label{lemma2_eqn1}
\end{align}
Since we are using uncoded QAM for transmission, the minimum norm
distance between any two elements $\mathbf{x}\neq\mathbf{x'} \in
\mathcal{X}^{(T_s-\nu)}$ is lower bounded by,
\begin{align*}
\|\mathbf{x}-\mathbf{x'}\|^2 & \stackrel{\cdot}{\geq}  SNR^{(1-r)}.
\end{align*}
From the fact that $\bfhat{Q}$ is full rank its smallest
singular value is nonzero and independent of SNR. Defining
$\bfhat{x}=\bfhat{Q} \mathbf{x}$ we can conclude that,
\begin{align}
\|\bfhat{x} - \bfhat{x}'\|^2 & \doteq   \|\mathbf{x}-\mathbf{x'}\|^2 
\stackrel{\cdot}{\geq} SNR^{(1-r)}.
\label{distance_preserved}
\end{align}
As $\bfhat{\Lambda}$ is a diagonal matrix,
\begin{align}
\|\bfhat{\Lambda} (\bfhat{x}-\bfhat{x}') \|^2 & = \sum_{l=0}^{T_s-\nu-1} | \lambda_l (\bfhat{x} - \bfhat{x}')_l |^2 =   \sum_{l=0}^{T_s-\nu-1} | \lambda_l |^2  | (\bfhat{x} - \bfhat{x}')_l |^2 
 \doteq  SNR^{-(1-r)+\epsilon} \sum_{l=0}^{T_s-\nu-1} | (\bfhat{x} - \bfhat{x}')_l |^2 \label{use_lemma}\\
& =   SNR^{-(1-r)+\epsilon} \| (\bfhat{x}-\bfhat{x}') \|^2 
 \stackrel{\cdot}{\geq}   SNR^{-(1-r)+\epsilon} SNR^{(1-r)} = SNR^{\epsilon} \nonumber
\end{align}
where (\ref{use_lemma}) is true from lemma \ref{lem:TapMag} for some
$\epsilon>0$. Since $Q(x)$ is a decreasing function in $x$, using the
above equation, we conclude that if
$\mathbf{h}\in\overline{\mathcal{M}(1-r)}$ the pairwise error
probability of detecting the sequence $\mathbf{x'}$ given that
$\mathbf{x}$ was transmitted is upper bounded by,
\begin{align}
P_e(\mathbf{x}\rightarrow \mathbf{x'}) & \leq  Q\left( \|\bfhat{\Lambda} (\bfhat{x}-\bfhat{x}') \|^2\right)  \stackrel{\cdot}{\leq} Q  \left( SNR^{\epsilon}   \right). \nonumber
\end{align}
Therefore, by the union bound we have,
\begin{align*}
P_e(SNR) & \stackrel{\cdot}{\leq}  SNR^r Q  \left( SNR^{\epsilon}  \right) \stackrel{\cdot}{\leq} SNR^r e^{-\frac{SNR^{2\epsilon}}{2}},
\end{align*}
as $Q(x)$ decays exponentially in $x$ for large $x$ {\em i.e.}, $Q(x)
 \leq e^{-\frac{x^2}{2}}$. Therefore we conclude that using the
 specific uncoded scheme described in Lemma \ref{lem:ExpDecOne}, if
 $\mathbf{h} \in \overline{\mathcal{M}(1-r)}$, the error probability
 $P_e$ decays exponentially in $SNR$.
\end{IEEEproof}

\subsection{Proof of Lemma \ref{lem:ExpDecTwo}}
\begin{IEEEproof}
Denote the transmitted sequence of length $T_s-\nu$ from the higher
and lower layer as $\mathbf{x_H}\in \mathcal{X}_H^{(T_s-\nu)}$ and
$\mathbf{x_L}\in \mathcal{X}_L^{(T_s-\nu)}$ respectively. For decoding
the higher layer we treat the signal on the lower layer as
noise. Proceed as in the proof of the Lemma \ref{lem:ExpDecOne} (\ref{lemma2_eqn1}) with the selection matrix $\mathbf{S}$ chosen such
that $\bfhat{\Lambda}=diag\left( \{\lambda_l:l\in
\mathcal{G}^{(1,1)}\}\right)$, where $|\mathcal{G}^{(1,1)}|\geq
(T_s-\nu)$. We get,
\begin{align*}
\bfhat{y} & =  \mathbf{S} \mathbf{\tilde{y}} = \bfhat{\Lambda} \underbrace{\bfhat{Q}\mathbf{x_H}}_{\bfhat{x}_H}+\bfhat{\Lambda}\underbrace{ \bfhat{Q}\mathbf{x_L}}_{\bfhat{x}_L} + \bfhat{z}  =   \bfhat{\Lambda}\bfhat{x}_H + \underbrace{\bfhat{\Lambda}\bfhat{x}_H+\bfhat{z}}_\mathbf{\tilde{z}} 
=  \bfhat{\Lambda}\bfhat{x}_H+ \mathbf{\tilde{z}}. 
\end{align*}
The decoding rule we use to decode $\mathbf{x}_H$ is given by,
\begin{align*}
\mathbf{\tilde{x}}_H & =  \argmin_{\mathbf{x}_H} \| \bfhat{y}-\bfhat{\Lambda}\bfhat{Q} \mathbf{x}_H \|^2.
\end{align*}
Therefore, the pairwise error probability of detecting the sequence
$\mathbf{x_H^{'}}$ if $\mathbf{x_H}$ was transmitted is given by,
\begin{align}
\lefteqn{
P_e^H(\mathbf{x}_H \rightarrow \mathbf{x}_H^{'})  =  \sum_{\mathbf{x}_L \in \mathcal{X}_L^{(T_s-\nu)} } Pr(\mathbf{x}_L) P_e(\mathbf{x_H}\rightarrow \mathbf{x}_H^{'} | \mathbf{\Lambda},\mathbf{x}_L ) }& \nonumber \\
& =  \sum_{\mathbf{x}_L \in \mathcal{X}_L^{(T_s-\nu)} } Pr(\mathbf{x}_L) Pr\left( \|\bfhat{y}-\bfhat{\Lambda} \bfhat{x}_H \|^2 > \|\bfhat{y}-\bfhat{\Lambda} \bfhat{x}_H^{'} \|^2 \right) \nonumber \\
& =  \sum_{ \mathbf{x}_L \in \mathcal{X}_L^{(T_s-\nu)}  } Pr(\mathbf{x}_L) Q \left( \| \bfhat{\Lambda}(\bfhat{x}_H-\bfhat{x}_H^{'})\| + 2 Re\frac{<\bfhat{\Lambda}(\bfhat{x}_H-\bfhat{x}_H^{'}),\bfhat{\Lambda}\bfhat{x}_L> }{\| \bfhat{\Lambda}(\bfhat{x}_H-\bfhat{x}_H^{'})\|} \right). \label{pep}
\end{align}
Note that $Q(x)$ is a decreasing function in $x$. Therefore, the
equation (\ref{pep}) is upper bounded by,
\begin{align}
P_e^H(\mathbf{x}_H \rightarrow \mathbf{x}_H^{'}) & \leq \sum_{ \mathbf{x}_L \in \mathcal{X}_L^{(T_s-\nu)}  } Pr(\mathbf{x}_L) Q \left(  \underbrace{\| \bfhat{\Lambda}(\bfhat{x}_H-\bfhat{x}_H^{'})\| - 2\|\bfhat{\Lambda}\bfhat{x}_L\|}_\Omega  \right) 
\end{align}
Define $\Gamma_{min}$ and $\Gamma_{max}$ as,
\begin{align*}
\Gamma_{min} =  \min_{i\in \mathcal{G}^{(1,1)}} |\lambda_i|^2, &\,\,
\Gamma_{max} =  \max_{i\in \mathcal{G}^{(1,1)}} |\lambda_i|^2.
\end{align*}
Therefore, from lemma \ref{lem:TapMag}, we get
\begin{align*}
\Gamma_{min} & \doteq  \Gamma_{max} \doteq  \max_{l\in\{ 0,1,\ldots,\nu\} }|h_l|^2 
 \doteq  SNR^{-(1-\tilde{r}_H)+ 2 \epsilon}
\end{align*}
where the last equality follows for some $\epsilon>0$ from lemma
\ref{lem:TapMag} as $\mathbf{h}\in \overline{\mathcal{M}(1-r_H)}$. Since
$\|\bfhat{x}_L\|^2 \stackrel{\cdot}{\leq} SNR^{1-\beta}$ and from
equation (\ref{distance_preserved}) in the proof of Lemma
\ref{lem:ExpDecOne}, we can lower bound $\Omega$ as,
\begin{align*}
\Omega & \geq  \Gamma_{min}^{\frac{1}{2}}  \| (\bfhat{x}_H-\bfhat{x}_H^{'})\| - 2 \Gamma_{max}^{\frac{1}{2}} \|\bfhat{x}_L\|
 \doteq  SNR^{\frac{-(1-\tilde{r}_H)+ 2 \epsilon}{2}}\left( \| \bfhat{x}_H-\bfhat{x}_H^{'}\| - \|\bfhat{x}_L\|\right) \\
& \doteq SNR^{-\frac{(1-\tilde{r}_H)}{2}+\epsilon}\left( SNR^{\frac{1-\tilde{r}_H}{2}}-SNR^{\frac{1-\beta}{2}}  \right) 
\doteq SNR^{\epsilon}, 
\end{align*}
where the last step is valid as $\beta>\tilde{r}_H$. Therefore 
$P_e^H (\mathbf{x}_H \rightarrow \mathbf{x}_H^{'}) \stackrel{\cdot}{\leq} Q(SNR^{\epsilon} )$, which 
decays exponentially in SNR. By the union bound as in the Lemma
\ref{lem:ExpDecOne} we conclude that given that
$\mathbf{h}\in\overline{\mathcal{M}(1-r_H)}$, $P_e^{H}(SNR)$ decays
exponentially in SNR even with superposition coding.
\end{IEEEproof}

\section{Proof of Lemmas for MISO ISI channel}
\subsection{Proof of Lemma \ref{lem:Decoder}}
\begin{IEEEproof}
Representing $\| \cdot \|$ to be the Frobenius norm and using the
decoding rule in (\ref{decoding_rule}), we get,
\begin{align}
\lefteqn{P_e(\Xbf_H\rightarrow \Xbf_H^{'} | \hbf ,\Xbf_L )  =  Pr\left( \|\bfhat{Y}- \bfhatLambda \bfbreveQ^{*} \Xbf_H \|^2 > \|\bfhat{Y}- \bfhatLambda \bfbreveQ^{*} \Xbf_H^{'} \|^2 \right)}& \nonumber \\
& =  Pr\left( \|\bfhatLambda \bfbreveQ^{*} \Xbf_L+\bfhat{Z} \|^2 > \|\bfhatLambda \bfbreveQ^{*} \Xbf_H - \bfhatLambda \bfbreveQ^{*} \Xbf_H^{'}+ \bfhatLambda \bfbreveQ^{*}\Xbf_L + \bfhat{Z} \|^2 \right). \label{decoder_eqn1} 
\end{align}
Denote $\xbf_L^{(i)},\zbf^{(i)},\xbf_H^{(i)},\ybf^{(i)}$ to be the
$i^{th}$ columns of $\Xbf_L$, $\bfhat{Z}$, $\Xbf_H$ and $\bfhat{Y}$
respectively. With these definitions we can expand the left hand side
and the right hand side of the inequality above as,
\begin{align*}
LHS & = \|\bfhatLambda \bfbreveQ^{*} \Xbf_L+\bfhat{Z} \|^2  = \sum_{i=1}^{T_b} \|\bfhatLambda \bfbreveQ^{*} \xbf_L^{(i)}+\zbf^{(i)} \|^2 \qquad \text{and}\\
RHS & = \|\bfhatLambda \bfbreveQ^{*} \Xbf_H - \bfhatLambda \bfbreveQ^{*} \Xbf_H^{'}+ \bfhatLambda \bfbreveQ^{*}\Xbf_L + \bfhat{Z} \|^2 
 = \sum_{i=1}^{T_b} \|\bfhatLambda \bfbreveQ^{*} \xbf_H^{(i)} - \bfhatLambda \bfbreveQ^{*} \xbf_H^{(i)'} + \bfhatLambda \bfbreveQ^{*}\xbf_L^{(i)} + \zbf^{(i)} \|^2 \\
& =  \sum_{i=1}^{T_b}  \left\{\|\bfhatLambda \bfbreveQ^{*} ( \xbf_H^{(i)} - \xbf_H^{(i)'})\|^2 + \|\bfhatLambda \bfbreveQ^{*}\xbf_L^{(i)} + \zbf^{(i)} \|^2 + 2 Re \left\langle \bfhatLambda \bfbreveQ^{*} ( \xbf_H^{(i)} - \xbf_H^{(i)'}),\bfhatLambda \bfbreveQ^{*}\xbf_L^{(i)} + \zbf^{(i)}\right\rangle \right\}.
%LHS & = \|\bfhatLambda \bfbreveQ^{*} \Xbf_L+\bfhat{Z} \|^2  = \sum_{i=1}^{T_b} \|\bfhatLambda \bfbreveQ^{*} \xbf_L^{(i)}+\zbf^{(i)} \|^2 \qquad \text{and}\\
%RHS & = \|\bfhatLambda \bfbreveQ^{*} \Xbf_H - \bfhatLambda \bfbreveQ^{*} \Xbf_H^{'}+ \bfhatLambda \bfbreveQ^{*}\Xbf_L + \bfhat{Z} \|^2 \\
%& = \sum_{i=1}^{T_b} \|\bfhatLambda \bfbreveQ^{*} \xbf_H^{(i)} - \bfhatLambda \bfbreveQ^{*} \xbf_H^{(i)'} + \bfhatLambda \bfbreveQ^{*}\xbf_L^{(i)} + \zbf^{(i)} \|^2 \\
%& =  \sum_{i=1}^{T_b}  \left\{\|\bfhatLambda \bfbreveQ^{*} ( \xbf_H^{(i)} - \xbf_H^{(i)'})\|^2 + \|\bfhatLambda \bfbreveQ^{*}\xbf_L^{(i)} + \zbf^{(i)} \|^2 \right.\\
%& \hspace{1in} \left.+ 2 Re \left\langle \bfhatLambda \bfbreveQ^{*} ( \xbf_H^{(i)} - \xbf_H^{(i)'}),\bfhatLambda \bfbreveQ^{*}\xbf_L^{(i)} + \zbf^{(i)}\right\rangle \right\}.
\end{align*}
Substituting these expansions in (\ref{decoder_eqn1}) and expanding we get,
\begin{align*}
\lefteqn{ \hspace{-0.6in}P_e(\Xbf_H\rightarrow \Xbf_H^{'} | \hbf ,\Xbf_L )    = 
Pr \left( - \sum_{i=1}^{T_b} \left( 2 Re \left\langle \bfhatLambda \bfbreveQ^{*} ( \xbf_H^{(i)} - \xbf_H^{(i)'}),\zbf^{(i)} \right \rangle\right) > 
\sum_{i=1}^{T_b} \left(  \|\bfhatLambda \bfbreveQ^{*} ( \xbf_H^{(i)} - \xbf_H^{(i)'})\|^2\right) +  \right. }& \\
& \hspace{2in} \left. \sum_{i=1}^{T_b} \left( 2 Re \left\langle \bfhatLambda \bfbreveQ^{*} ( \xbf_H^{(i)} - \xbf_H^{(i)'}),\bfhatLambda \bfbreveQ^{*}\xbf_L^{(i)} \right\rangle  \right) \right).
%\lefteqn{ \hspace{-0.6in}P_e(\Xbf_H\rightarrow \Xbf_H^{'} | \hbf ,\Xbf_L )    = Pr \left( \sum_{i=1}^{T_b} \|\bfhatLambda \bfbreveQ^{*} \xbf_L^{(i)}+\zbf^{(i)} \|^2  >  \sum_{i=1}^{T_b} \|\bfhatLambda \bfbreveQ^{*} \xbf_H^{(i)} - \bfhatLambda \bfbreveQ^{*} \xbf_H^{(i)'} + %\bfhatLambda \bfbreveQ^{*}\xbf_L^{(i)} + \zbf^{(i)} \|^2\right) }& \\
%\lefteqn{ P_e(\Xbf_H\rightarrow \Xbf_H^{'} | \hbf ,\Xbf_L )    = Pr \left( \sum_{i=1}^{T_b} \|\bfhatLambda \bfbreveQ^{*} \xbf_L^{(i)}+\zbf^{(i)} \|^2  > \right.}&\\
%& \hspace{1.5in}\left. \sum_{i=1}^{T_b} \|\bfhatLambda \bfbreveQ^{*} \xbf_H^{(i)} - \bfhatLambda \bfbreveQ^{*} \xbf_H^{(i)'} + \bfhatLambda \bfbreveQ^{*}\xbf_L^{(i)} + \zbf^{(i)} \|^2\right)& \\
%& = Pr \left( 0  > \sum_{i=1}^{T_b} \left(  \|\bfhatLambda \bfbreveQ^{*} ( \xbf_H^{(i)} - \xbf_H^{(i)'})\|^2 + 2 Re \left\langle\bfhatLambda \bfbreveQ^{*} ( \xbf_H^{(i)} - \xbf_H^{(i)'}),\bfhatLambda \bfbreveQ^{*}\xbf_L^{(i)} + \zbf^{(i)}\right\rangle  \right) \right)  \\
%& = Pr \left( - \sum_{i=1}^{T_b} \left( 2 Re \left\langle \bfhatLambda \bfbreveQ^{*} ( \xbf_H^{(i)} - \xbf_H^{(i)'}),\zbf^{(i)} \right \rangle\right) > 
%\sum_{i=1}^{T_b} \left(  \|\bfhatLambda \bfbreveQ^{*} ( \xbf_H^{(i)} - \xbf_H^{(i)'})\|^2\right) +  \right. \\
%& \hspace{2in} \left. \sum_{i=1}^{T_b} \left( 2 Re \left\langle \bfhatLambda \bfbreveQ^{*} ( \xbf_H^{(i)} - \xbf_H^{(i)'}),\bfhatLambda \bfbreveQ^{*}\xbf_L^{(i)} \right\rangle  \right) \right)  
\end{align*}
Defining,
\begin{align*}
\mathbf{u}^{(i)} & = \frac{ \bfhatLambda \bfbreveQ^{*} ( \xbf_H^{(i)} - \xbf_H^{(i)'})}{ \sqrt{ \sum_{i=1}^{T_b} 
\|\bfhatLambda \bfbreveQ^{*} ( \xbf_H^{(i)} - \xbf_H^{(i)'})\|^2 } }
\end{align*}
we can see that, $\sum_{i=1}^{T_b} \mathbf{u}^{(i) * } \mathbf{u}^{(i)}  = \sum_{i=1}^{T_b} \|\mathbf{u}^{(i)}\|^2 = 1$. Dividing both sides by $\sqrt{ \sum_{i=1}^{T_b} \|\bfhatLambda
  \bfbreveQ^{*} ( \xbf_H^{(i)} - \xbf_H^{(i)'})\|^2 } $ we get,
\begin{align*}
P_e(\Xbf_H\rightarrow \Xbf_H^{'} | \hbf ,\Xbf_L )  & = 
Pr \left( v > \sqrt{ \sum_{i=1}^{T_b} \left(  \|\bfhatLambda \bfbreveQ^{*} ( \xbf_H^{(i)} - \xbf_H^{(i)'})\|\right)} + \sum_{i=1}^{T_b} \left( 2 Re \left\langle \mathbf{u}^{(i)},\bfhatLambda \bfbreveQ^{*}\xbf_L^{(i)}\right\rangle  \right) \right)
%P_e(\Xbf_H\rightarrow \Xbf_H^{'} | \hbf ,\Xbf_L )  & = 
%Pr \left( v > \sqrt{ \sum_{i=1}^{T_b} \left(  \|\bfhatLambda \bfbreveQ^{*} ( \xbf_H^{(i)} - \xbf_H^{(i)'})\|\right)} + \right.\\
%& \hspace{2in} \left. \sum_{i=1}^{T_b} \left( 2 Re \left\langle \mathbf{u}^{(i)},\bfhatLambda \bfbreveQ^{*}\xbf_L^{(i)}\right\rangle  \right) \right)
\end{align*}
where,
\begin{align*}
v &= \sum_{i=1}^{T_b} \left( 2 Re \left\langle \mathbf{u}^{(i)},-\zbf^{(i)} \right \rangle \right)  = \mathcal{C} \mathcal{N} \left( 0,\mathbb{E} \left( \sum_{i=1}^{T_b}  \mathbf{u}^{(i) * } \mathbf{u}^{(i)} \right) \right) = \mathcal{C} \mathcal{N} \left( 0 , 1 \right).
\end{align*}
Therefore,
\begin{align*}
 P_e(\Xbf_H\rightarrow \Xbf_H^{'} | \hbf ,\Xbf_L ) & = Q \left( \sqrt{  \sum_{i=1}^{T_b} \left(  \|\bfhatLambda \bfbreveQ^{*} ( \xbf_H^{(i)} - \xbf_H^{(i)'})\|^2\right)} + \sum_{i=1}^{T_b} \left( 2 Re \left\langle \mathbf{u}^{(i)},\bfhatLambda \bfbreveQ^{*}\xbf_L^{(i)} \right\rangle  \right) \right) \\
& \leq Q \left( \sqrt{  \sum_{i=1}^{T_b} \left(  \|\bfhatLambda \bfbreveQ^{*} ( \xbf_H^{(i)} - \xbf_H^{(i)'})\|^2\right)}  - 2 \sum_{i=1}^{T_b} \left( \| \mathbf{u}^{(i)}\|. \| \bfhatLambda \bfbreveQ^{*}\xbf_L^{(i)} \|  \right) \right) \\
& \leq Q \left(  \sqrt{ \sum_{i=1}^{T_b} \left(  \|\bfhatLambda \bfbreveQ^{*} ( \xbf_H^{(i)} - \xbf_H^{(i)'})\|^2\right)}  - 2 \sum_{i=1}^{T_b} \left( \| \bfhatLambda \bfbreveQ^{*}\xbf_L^{(i)} \|  \right) \right).
\end{align*}
Since $\sum_{i=1}^{T_b} \|\bfhatLambda  \bfbreveQ^{*} ( \xbf_H^{(i)} - \xbf_H^{(i)'})\|^2  = \|\bfhatLambda  \bfbreveQ^{*} ( \Xbf_H - \Xbf_H^{'})\|^2 $ we can rewrite the equation to get the desired result {\em i.e.},
\begin{align*}
P_e(\Xbf_H\rightarrow \Xbf_H^{'} | \hbf ,\Xbf_L ) & \leq Q \left(  \|\bfhatLambda  \bfbreveQ^{*} ( \Xbf_H - \Xbf_H^{'})\|  - 2 \sum_{i=1}^{T_b} \left( \| \bfhatLambda \bfbreveQ^{*}\xbf_L^{(i)} \|  \right) \right).
\end{align*}
\end{IEEEproof}

\subsection{Proof of Lemma \ref{SVDQDeltaX}}

\begin{IEEEproof}
Since $\bfbreveQ^* \Delta \Xbf_H \Delta \Xbf_H^{*} \bfbreveQ$ is a
Hermitian matrix it can be written as in (\ref{SVD_second}).  Since
$\bfhat{Q}^*$ is still a full rank Vandermonde matrix which does not
depend on SNR it follows that is a full rank matrix independent of
$SNR$. Since determinant of a matrix is product of its eigenvalues we
get,
\begin{align*}
  \prod_{i=1}^{(T_s-\nu)M_t} \xi_i^2 &= det( \bfbreveQ^* \Delta \Xbf_H  \Delta \Xbf_H^{*} \bfbreveQ )
 = det( \bfbreveQ \bfbreveQ^* \Delta \Xbf_H  \Delta \Xbf_H^{*} )  
 =  det( \bfbreveQ \bfbreveQ^*)det(\Delta \Xbf_H  \Delta \Xbf_H^{*} )  \nonumber \\
  & \doteq det(\Delta \Xbf_H  \Delta \Xbf_H^{*} ) \stackrel{\cdot}{\geq} SNR^{(T_s-\nu)M_t-(T_s-\nu)r} \qquad \text{from (\ref{constraint3})}.
\end{align*}
Combining submultiplicativity of the Frobenius norm\footnote{$\|AB\|_F\leq \|A\|_F \|B\|_F $ }  with the fact that
the sum of the eigenvalues is equal to the trace, for all $i \in
\{1,\ldots,(T_s-\nu)M_t\}$ we get, 
\begin{align*}
  \xi_i^2 & \leq \sum_{i=1}^{(T_s-\nu)M_t} \xi_i^2  =  tr \left( \bfbreveQ^* \Delta \Xbf_H  \Delta \Xbf_H^{*} \bfbreveQ \right) 
 = \| \bfbreveQ^* \Delta \Xbf_H \|_F^2  
 \leq \| \bfbreveQ^* \|_F^2 \| \Delta \Xbf_H \|_F^2 \doteq SNR \quad \text{from  (\ref{constraint4})}. 
\end{align*}
\end{IEEEproof}

\subsection{Proof of Lemma \ref{SVDLambda}}
%Beginning of proof of LEMMA on SVD OF \bfhatLambda^* \bfhatLambda 
\begin{IEEEproof}
Observe that because of the way we have chosen our selection matrix,
\begin{align*}
  \bfhat{\Lambda} \bfhat{\Lambda}^{*} & =  \displaystyle {\rm diag}_{i\in \mathcal{G}^{(1,\kappa)} } \big(\big\{ \sum_{p=1}^{M_t} |\lambda_i^{(1,p)}|^2 \big\} \big) 
 \doteq \displaystyle {\rm diag}_{i \in \mathcal{G}^{(1,\kappa)}} \big( \big\{ |\lambda_i^{(1,\kappa)}|^2 \big\} \big) 
\end{align*}
as $|\lambda_i^{(1,\kappa)}|^2$ is the dominant term in the
summation. From Lemma \ref{lem:TapMag} we have that for $i \in
\mathcal{G}^{(1,\kappa)}$, 
%\begin{align*}
$|\lambda_i^{(\kappa)}|^2  \doteq\max_{i\in\{0,1,\ldots,\nu\} } |h_i^{(1,\kappa)}|^2.$
%\end{align*}
We know that the eigenvalues of $\bfhat{\Lambda} \bfhat{\Lambda}^{*}$
and $\bfhatLambda^* \bfhatLambda $ are identical with the remaining
eigenvalues being equal to zero. Assume that $\mathbf{D}_3^2$ is
represented as, $\mathbf{D}_3^2 = diag \left( \gamma_1^2,
\gamma_2^2,\ldots,\gamma_{(T_s-\nu)}^2,\ldots,\gamma_{(T_s-\nu)M_t}^2 \right)$ where
$\gamma_1^2 \geq \gamma_2^2 \ldots \geq \gamma_{(T_s-\nu) M_t}^2$. The result
then follows directly.
\end{IEEEproof}

\subsection{Proof of Lemma \ref{lem:MISOEXPDecay}}
\begin{IEEEproof}
For a uniform choice of codewords for the lower layer from lemma
\ref{lem:Decoder}, using the decoding rule in (\ref{decoding_rule}),
the pairwise error probability of detecting the sequence $\Xbf_H^{'}$
given that $\Xbf_H$ was transmitted is upper bounded by,
\begin{align}
\lefteqn{ P_e(\Xbf \rightarrow \Xbf^{'}|\hbf \in \overline{\mathcal{M}(1-r_H)})  =  \sum_{\Xbf_L } Pr(\Xbf_L) P_e(\Xbf_H\rightarrow \Xbf_H^{'} | \Xbf_L,\hbf \in \overline{\mathcal{M}(1-r_H)} )}&  \nonumber \\
&  =   SNR^{-r_L T}\sum_{\Xbf_L } P_e(\Xbf_H\rightarrow \Xbf_H^{'} | \Xbf_L,\hbf \in \overline{\mathcal{M}(1-r_H)} ) 
%\nonumber \\
%& 
\leq Q \bigg( \underbrace{  \| \bfhatLambda \bfbreveQ^{*} (\Xbf_H-\Xbf_H^{'})\| - 2\sum_{i=1}^{T_b} \left( \| \bfhatLambda \bfbreveQ^{*}\xbf_L^{(i)} \|  \right) }_\Omega  \bigg) \nonumber
\end{align}
where $\xbf_L^{(i)}$ is the $i^{th}$ column of $\Xbf_L$. We will now
get a lower bound on $\Omega$ in the equation above to get an
upper bound to the error probability. Using $tr(AB)=tr(BA)$ and
representing $\Delta \Xbf_H=\Xbf_H-\Xbf_H^{'}$, for the first term in
$\Omega$ we get that,
\begin{align}
  \| \bfhatLambda \bfbreveQ^{*} (\Xbf_H-\Xbf_H^{'})\|^2 & = tr\left( \bfhatLambda \bfbreveQ^{*} \Delta \Xbf_H \Delta \Xbf_H^{*} \bfbreveQ \bfhatLambda^* \right)  
 = tr\left(  \bfbreveQ^{*} \Delta \Xbf_H  \Delta \Xbf_H^{*} \bfbreveQ \bfhatLambda^* \bfhatLambda \right)  \label{term1_1}
\end{align}
Substituting the SVD from (\ref{SVD_second}) and Lemma \ref{SVDLambda}
into (\ref{term1_1}) we get,
\begin{align}
  \| \bfhatLambda \bfbreveQ^{*} (\Xbf_H-\Xbf_H^{'})\|^2 & = tr\left( \mathbf{R} \mathbf{D}_2^2 \mathbf{R}^*  \mathbf{V}^* \mathbf{D}_3^2 \mathbf{V} \right) 
 =  tr\left( \mathbf{V} \mathbf{R} \mathbf{D}_2^2 \mathbf{R}^*  \mathbf{V}^* \mathbf{D}_3^2  \right) 
 =  tr\left( \mathbf{T} \mathbf{D}_2^2 \mathbf{T}^*  \mathbf{D}_3^2  \right)   \nonumber \\
& =  \sum_{i,j=1}^{(T_s-\nu) M_t} \gamma_i^2 \xi_j^2 |t_{ij}|^2 \label{term1_2} 
\end{align}
where $\mathbf{T}=\mathbf{V}\mathbf{R}$ is also an unitary matrix and
$t_{ij}$ is the $(i,j)$ element of $\mathbf{T}$. Since,
\begin{align*}
\xi_1 \leq \xi_2 \leq \ldots \leq \xi_{ (T_s-\nu)M_t}, & \qquad \gamma_1^2 \geq \gamma_2^2 \ldots \geq \gamma_{(T_s-\nu)M_t}^2
\end{align*}
using similar reasoning as \cite{KosWes03,EliKum06} in (\ref{term1_2}) we get,
\begin{align}
\lefteqn{\| \bfhatLambda \bfbreveQ^{*} (\Xbf_H-\Xbf_H^{'})\|^2  \geq  \sum_{i=1}^{(T_s-\nu)M_t} \gamma_i^2 \xi_i^2 
\stackrel{(a)}{=} \sum_{i=1}^{(T_s-\nu)} \gamma_i ^2 \xi_i^2 
 \doteq \sum_{i=1}^{(T_s-\nu)} \lambda^2 \xi_i^2  \stackrel{(b)}{\geq}  \lambda^2 (T_s-\nu) \left[ \prod_{i=1}^{(T_s-\nu)} \xi_i^2 \right]^{\frac{1}{(T_s-\nu)}} } & \nonumber \\
& \stackrel{(c)}{\stackrel{\cdot}{\geq}}  \lambda^2 \left[ \frac{SNR^{(T_s-\nu)M_t-(T_s-\nu)r_H}}{\prod_{i=(T_s-\nu)+1}^{(T_s-\nu)M_t} \xi_i^2} \right]^{\frac{1}{(T_s-\nu)}} \geq  \lambda^2 \left[ \frac{SNR^{(T_s-\nu)M_t-(T_s-\nu)r_H}}{\prod_{i=(T_s-\nu)+1}^{(T_s-\nu)M_t} \xi_{max}^2} \right]^{\frac{1}{(T_s-\nu)}} \nonumber \\
& \stackrel{(d)}{ \stackrel{\cdot}{\geq}} \lambda^2 \left[ \frac{SNR^{(T_s-\nu)M_t-(T_s-\nu)r_H}}{ SNR^{(T_s-\nu)M_t-(T_s-\nu)} } \right]^{\frac{1}{(T_s-\nu)}} 
= \lambda^2 \left[ SNR^{(T_s-\nu)(1-r_H)} \right]^{\frac{1}{(T_s-\nu)}} 
 = \lambda^2 SNR^{(1-r_H)}. \nonumber 
\end{align}
where $(a)$ follows from(\ref{value_gamma}), $(b)$ follows from AM$\geq$GM, $(c)$ follows from (\ref{lower_bound_product}) and $(d)$ is from (\ref{upper_bound_max}). Given that $\mathbf{h} \in \overline{\mathcal{M}(1-r_H)}$ we can write,
\begin{align}
\lambda^2  & \doteq |h_l^{(1,\kappa)}|^2 \doteq SNR^{2\epsilon}  SNR^{-(1-r_H)}  \label{def_lambda}
\end{align}
where $\epsilon>0$. Therefore,
\begin{align}
\| \bfhatLambda \bfbreveQ^{*} (\Xbf_H-\Xbf_H^{'})\|^2 & \stackrel{\cdot}{\geq} SNR^{2\epsilon} SNR^{-(1-r_H)} SNR^{(1-r_H)}  = SNR^{2\epsilon}. \label{term1}
\end{align}
For the second term in $\Omega$, from the submultiplicativity of the
Frobenius Norm we get
\begin{align}
 \|\bfhatLambda \bfbreveQ^{*} \xbf_L^{(i)}\|^2  & \leq  \|\bfhatLambda\|^2 \|\bfbreveQ^{*}\|^2  \|\xbf_L^{(i)}\|^2 \doteq \|\bfhatLambda\|^2   \|\xbf_L^{(i)}\|^2 \doteq \lambda^2 \|\xbf_L^{(i)}\|^2   
\stackrel{\cdot}{\leq}\lambda^2 \|\Xbf_L\|^2 \nonumber \\
&\stackrel{(a)}{ \stackrel{\cdot}{\leq}} SNR^{2\epsilon} SNR^{-(1-r_H)} SNR^{1-\beta} \doteq SNR^{2\epsilon} SNR^{r_H-\beta} \label{term2} 
\end{align}
where $(a)$ follows from (\ref{constraint2_lower}) and (\ref{def_lambda}). Therefore, combining equations (\ref{term1}) and (\ref{term2}) we can
lower bound $\Omega$ as,
\begin{align*}
\Omega &\stackrel{\cdot}{\geq} SNR^{\epsilon} - T_b SNR^{\epsilon + \frac{(r_H-\beta)}{2}}  
 \doteq SNR^{\epsilon}\left( 1 - T_b SNR^{\frac{(r_H-\beta)}{2}}\right)
 \doteq SNR^{\epsilon},
\end{align*}
where the last step is valid as $\beta>r_H$. Therefore,
\begin{align}
P_e^H(\Xbf_H \rightarrow \Xbf_H^{'}) & \stackrel{\cdot}{\leq} Q \left( SNR^{\epsilon}  \right) .
\end{align}
Note that $Q(x)$ decays exponentially in $x$ for large $x$ {\em i.e.},
$Q(x) \leq e^{-\frac{x^2}{2}}$. By the union bound it then follows
that given that $\hbf \in \overline{\mathcal{M}(1-r_H)}$, $P_e^H(SNR)$ decays
exponentially in SNR.  From the union bound and the exponential decay
of $Q(x)$ it then follows that given that $\hbf \in
\overline{\mathcal{M}(1-r_H)}$, $P_e^H(SNR)$ decays exponentially in
SNR.
\end{IEEEproof}
\end{appendix}

\bibliographystyle{IEEEtranS}
%\bibliography{references}
\newpage

\end{document}